\pdfoutput=1
\documentclass[cis]{ipart_v1}

\firstpage{255}
\Vol{16}
\Issue{4}
\Year{2016}

\usepackage[english]{babel}

\usepackage{tikz}
\usetikzlibrary{matrix,decorations.pathreplacing}

\usepackage{epsfig}
\usepackage{graphicx}
\usepackage{amsmath}
\usepackage{amssymb}

\usepackage{algorithm}
\usepackage{algorithmicx}
\usepackage{algpseudocode}
\usepackage{graphicx}
\usepackage{eso-pic}
\usepackage{xspace}
\usepackage{multirow}
\usepackage{color}

\usepackage{hyperref}
\hypersetup{pdfauthor={Fang},pdftitle={Automatic thread painting generation}}

\theoremstyle{plain}

\theoremstyle{definition}

\theoremstyle{definition}

\begin{document}

\title[Automatic thread painting generation]{Automatic thread painting generation}

\author[X.~Fang, B.~Liu, and A.~Shamir]{Xiao-Nan Fang, Bin Liu, and Ariel Shamir$^\ast$ \blfootnote{$^\ast$This work was supported by the National Key Technology R\&D Program(Project Number 2016YFB1001402) and the Joint NSFC-ISF Research Program (project number 61561146393).}
}

\begin{abstract}
ThreadTone is an NPR representation of an input image by half-toning using threads on a circle. Current approaches to create ThreadTone paintings greedily draw the chords on the circle. We introduce the concept of chord space, and design a new algorithm to improve the quality of the thread painting. We use an optimization process that estimates the fitness of every chord in the chord space, and an error-diffusion based sampling process that selects a moderate number of chords to produce the output painting. We used an image similarity measure to evaluate the quality of our thread painting and also conducted a user study. Our approach can produce high quality results on portraits, sketches as well as cartoon pictures.
\end{abstract}

\maketitle

\section{Introduction}

Non-photorealistic rendering (NPR) aims at creating various styles of digital art. Different from photorealism, NPR generates artistic renderings in various styles such as impressionist~\cite{litwinowicz1997processing}, watercolor~\cite{curtis1997computer}, pen-and-ink~\cite{salisbury1997orientable}, line-art~\cite{kang2005interactive} and animation~\cite{meier1996painterly}. Such techniques can be utilized to render 3D scenes~\cite{rossl2000line}, to change the style of an image~\cite{gatys2016image}, or to provide users an interactive tool for digital painting~\cite{haberli1990paint}.

\begin{figure}[tbp]
    \centering

        \begin{tabular}{c@{\hspace{1mm}}c@{\hspace{1mm}}c@{\hspace{1mm}}c@{\hspace{1mm}}c}
            \includegraphics[height=1.1in]{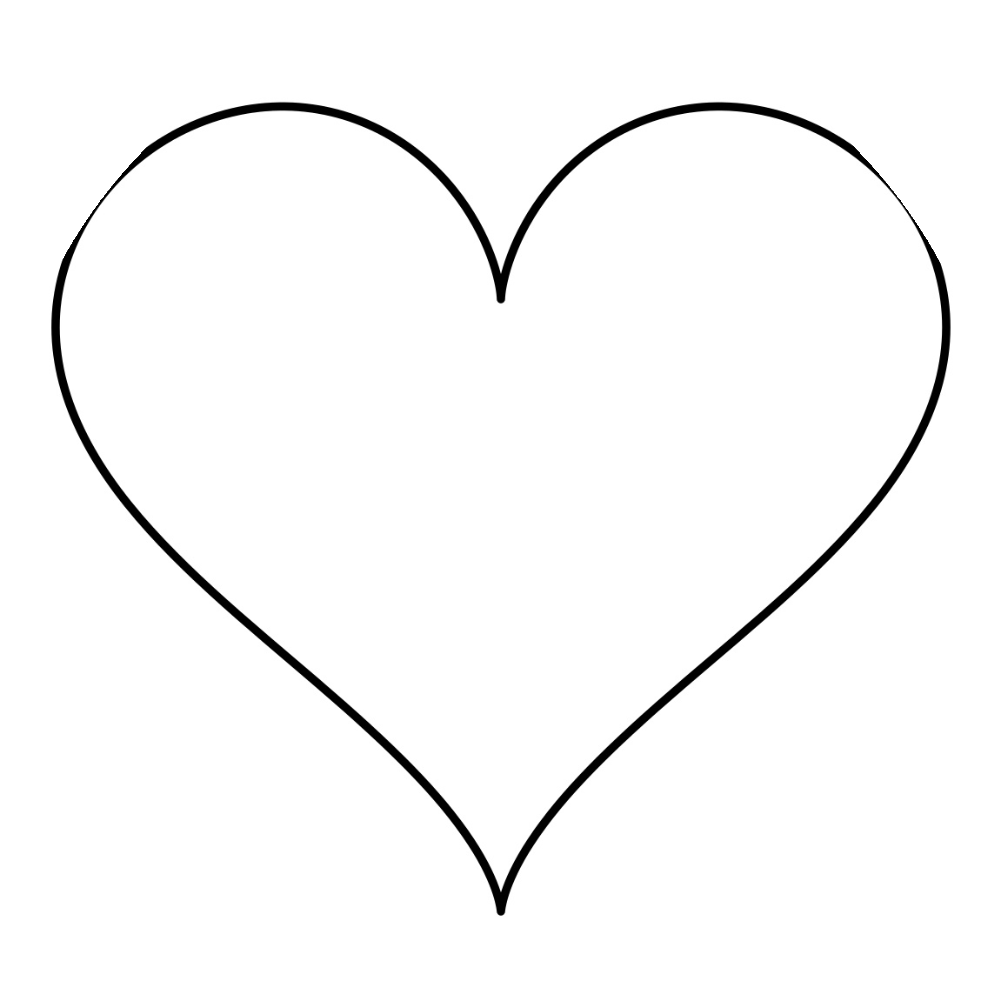}&
            \includegraphics[height=1.1in]{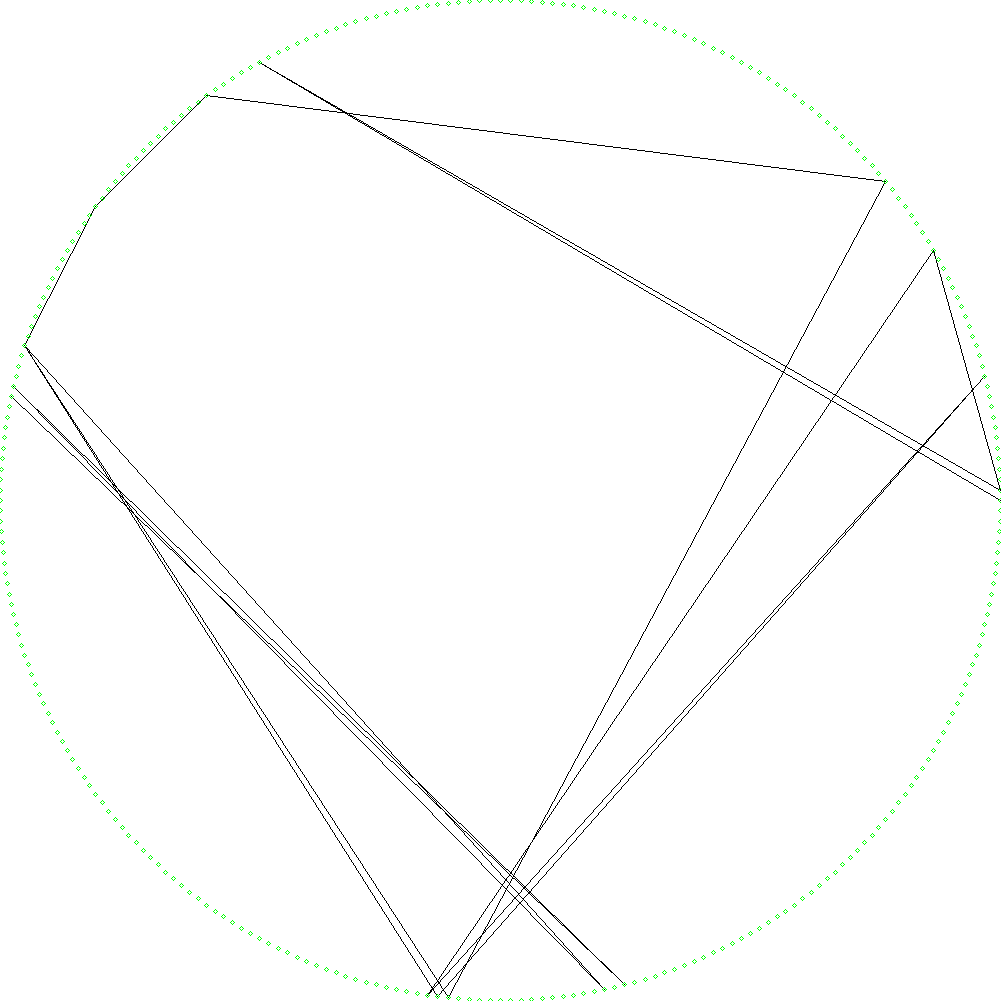}&
            \includegraphics[height=1.1in]{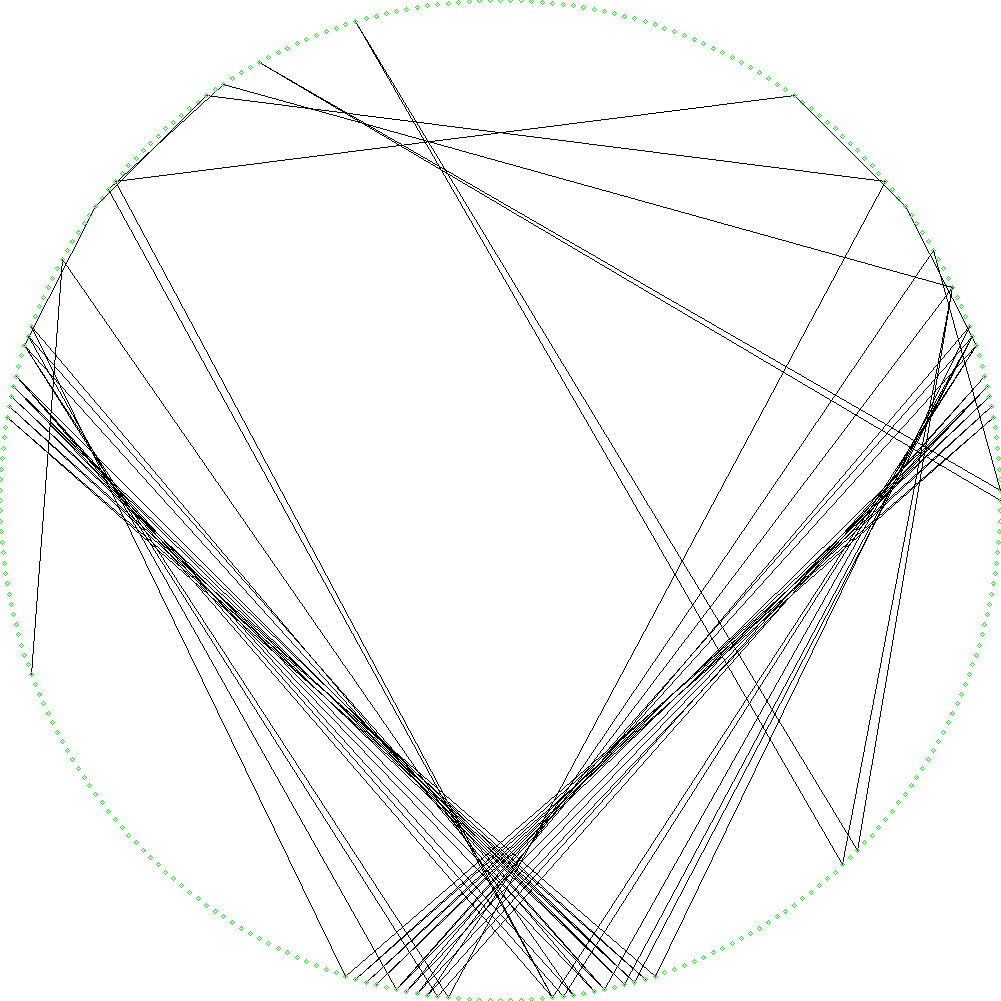}&
            \includegraphics[height=1.1in]{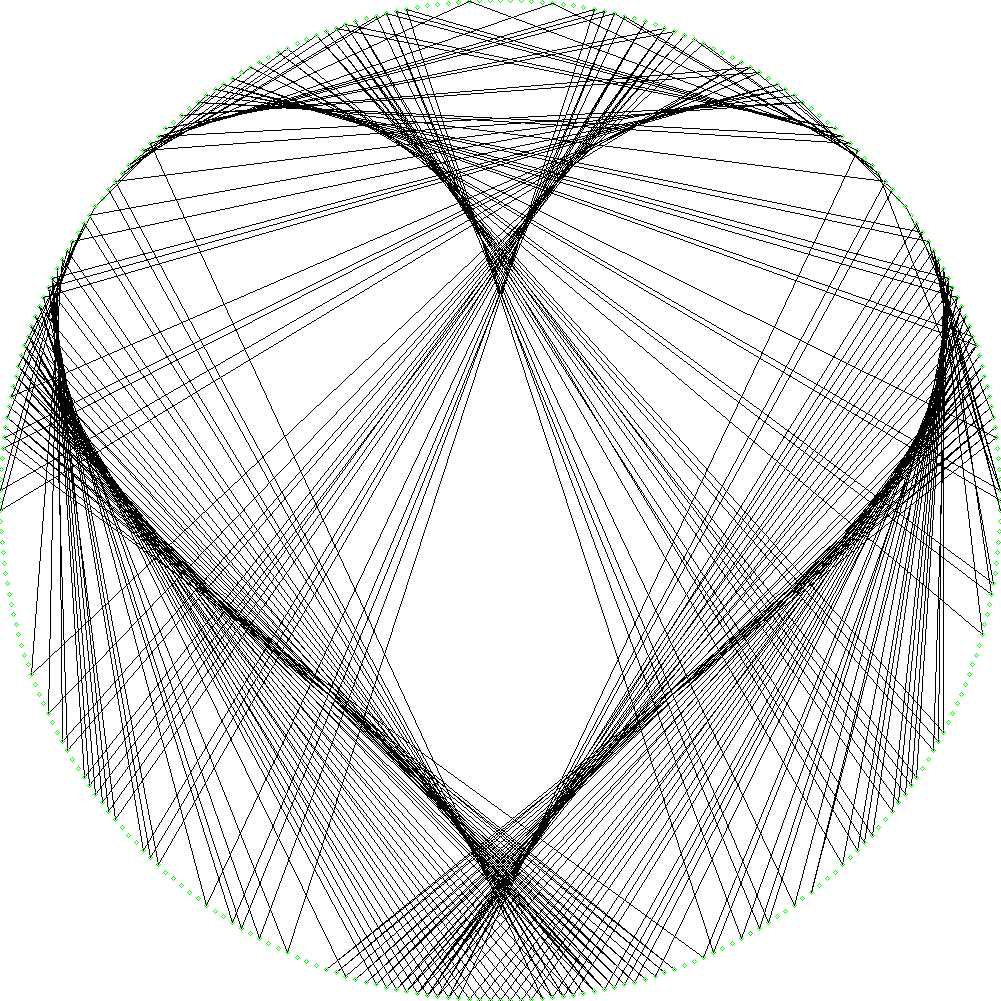}\\
            (a) Original image & (b) 16 lines & (c) 64 lines & (d) 256 lines 
        \end{tabular}
    \caption{Illustration of the thread painting process. The algorithm draws connected chords between the pins on the border of the circle. (a): The original image; (b)-(d): The drawing process, the number of chords increases from 16, 64 to 256.}
    \label{fig:description}
\end{figure}

Previous NPR algorithms usually produce an output picture as a collection of basic elements. Such elements include, for example, brush strokes~\cite{hertzmann1998painterly}, streamlines~\cite{hays2004image} or stipples~\cite{deussen2000floating}. Recently, Petros Vrellis, a new-media artist, introduced ThreadTone~\cite{vrellis} as a special type of line art.  In ThreadTone, the creator puts a number of pins on the circumference of a circular board, and winds a long thread through the pins to approximate an image of some object (see Fig.\ref{fig:description}). Our goal is to provide an automatic implementation for this kind of art with high quality results. 

Suppose that there are $P$ pins ${p_0, p_1,\dots,p_{P-1}}$ uniformly placed on the circumference of the circle, and that $k$ chord lines are used for the painting, then given an image $I$, the complete process of the thread painting generation involves the following steps:
\begin{enumerate}
\item Crop a circular region from the input image $I$.
\item Compute an appropriate ordered sequence of pins $\lbrace p_{i_0}, p_{i_1}, p_{i_2}, \dots, p_{i_k}\rbrace$ to cross with the thread.
\item Draw the line chords $\overline{p_{i_0}p_{i_1}},\overline{p_{i_1}p_{i_2}},\dots,\overline{p_{i_{k-1}}p_{i_k}}$ inside the circle (or wind a thread through the sequence of pins).
\end{enumerate}

The strategy of cropping a region of interest can dramatically effect the output result. Automatically choosing the region of interest is a challenging issue. In this work, we leave this task to the user. If a circular region is not specified, we simply crop the maximal circle region at the middle of input image $I$. The main challenge in this process lies in the second step: how to compute the sequence of pins so that the chords connecting them will approximate the appearance of the cropped circular region of the image $I$?  The requirement that the end of each chord is the beginning of the next one is called the \emph{connectivity requirement} and means that the painting can be created by winding a single long thread on the pins.

The most significant difference between this artistic style and previous ones is that the basic elements in a ThreadTone painting, namely the chords, are long straight lines (they may be as long as the diameter of the circle) that cannot bend or twist, while the strokes of other styles are flexible and have a limited size. Therefore, the properties of strokes or dots can be determined only within a small area of the input image, while to form chords we have to consider more global information. For example, if a chord crosses both a dark region and a bright region, it is not simple to judge whether it is suitable for drawing. Moreover, a single pixel in an image can be covered by many chords, which means that the number of chord intersections can be large, so the interrelations among chords must be carefully dealt with by the algorithm.

Our algorithm treats the drawing generation as a two-stage process: chord fitness computation and chord sampling. In the first stage, we solve a quadratic optimization problem to get the fitness value for every possible chord. A linear combination of the fitness values of all chords should recover the original image as accurately as possible. In the second stage, we calculate the sequence of chords by sampling $k$ chords with high fitness value, utilizing error diffusion to simulate the mutual impact between neighboring chords. The number of drawn chords $k$ could be either automatically determined or provided by the user.

We evaluated the performance of our algorithm by calculating SSIM index~\cite{wang2004image} between the image and the painting, and performing a user study. Our approach produces high-quality thread paintings and could reproduce the details of input objects although the form and element of such ThreadTone painting are strictly limited.

\section{Related works}

NPR covers many artistic styles. Haeberli presented a tool to generate an abstract image representation~\cite{haberli1990paint}. Users could draw strokes on a canvas to obtain an impressionist style of the source image. The brush-stroke representation was then extended for impressionist video synthesis by Litwinowicz~\cite{litwinowicz1997processing}. Salisbury {\em et al.} focused on pen-and-ink illustrations~\cite{salisbury1997orientable}. Deussen {\em et al.} represented images with a collection of stipples~\cite{deussen2000floating}. In particular, many efforts have been devoted to the generation of line representations for images. Kang {\em et al.} used streamlines to illustrate the vector field on images~\cite{kang2009flow}. Algorithms were proposed to generate line drawings from input images~\cite{kang2007coherent} or smooth surfaces~\cite{hertzmann2000illustrating,elber1999interactive}, and to mimic a specific artist line-drawing style~\cite{berger2013}. 
Style transfer algorithm by Hertzmann {\em et al.}~\cite{hertzmann2001image} searches for similar patches from the reference pair to fill a target image. Such algorithm could be applied on a large variety of tasks.
Recently, Chen {\em et al.}~\cite{chen2012embroidery} and Yang {\em et al.}~\cite{yang2016paint} made contribution to modeling embroidery, a traditional Chinese art style, that synthesized the picture with stitches.

One topic of NPR is to produce an output picture as a collection of basic elements.
The approaches for obtaining a suitable collection of basic elements can generally be divided into three types: greedy method, trial-and-error method and Voronoi-based method~\cite{hertzmann2003a}. The greedy method collects the basic elements one by one.  In every step, the element that has the maximum value of some heuristic function is selected~\cite{kolpatzik1992optimized}. The trial-and-error method, also called relaxation, is a randomized optimization method. It iteratively generates some small change suggested by the algorithm of the current state, and this new placement is accepted if it reduces some total measured energy. This method was adopted by~\cite{pang2008structure,haberli1990paint}. The Voronoi-based method forces the elements to be evenly placed on the canvas. The dot size or dot spacing could be adjusted according to the gray level of the original pixels~\cite{secord2002weighted}. The placement of elements is optimized via an iterative update process similar to the K-means clustering. 
Our work also produces a picture using a collection of chords. We first use a global optimization to get the weight for each possible chord element. Then, we use error-diffusion-based method to sample the elements, which will be discussed in Section~\ref{section:line_sampling}.


A recent algorithm to create ThreadTone painting from a given grayscale image $I$ was proposed in~\cite{ttweb}. The algorithm selects chords that fulfill the connectivity requirement to form the painting in a greedy manner. For convenience, the pixel values are reversed as $I_i \leftarrow (255 - I_i)$, so that larger values represent higher priority for drawing.

The first pin is chosen at random or by the user. Next, given the current pin $p_{curr}$ in the sequence of pins, the next pin $p_{next}$ is chosen as the pin whose chord $\overline{p_{curr}p_{next}}$ has maximum covering of pixel values, from all chords originating at $p_{curr}$ and ending at all possible pins. The covering of a chord is the sum of the values of the pixels it covers. To simulate the effect of the previously drawn chords, the value of each pixel covered by a chord is reduced by a certain amount (15 in their implementation). To avoid drawing chords among several pins back and forth, small loops are excluded from the possible selection.

Although this greedy method could achieve reasonable result, it cannot handle difficult cases and the drawbacks are obvious. First, the selection of chords lacks global control. Therefore, the algorithm tends to extensively draw chords over dark areas and to leave other area unpainted. Second, the reduction of pixel values brings dramatic change to the original image. As the number of chords increases, the structure of the main object will be destroyed. Moreover, the importance of pixels in the input image whose values are the same are usually not identical. For example, viewers usually focus on the eyes, nose and mouth of a portrait, while focusing less on the  background. Hence, the accuracy of covering the face is more important than the background. Our approach solves these problems and provides users with more freedom to adjust the sharpness of chords and the region of interest. In all our results we compare our approach to this baseline greedy algorithm.

\section{Method}

Given the image $I$, if it is not a grayscale image, we use the method proposed in~\cite{lu2012real} to convert it from color to grayscale.
Next, we either crop it to a circle denoted by the user or find the maximum circular region whose center is at the center of the image.
Given the number of chords $k$, instead of choosing the chords one by one greedily, we formulate the problem as a global optimization problem of choosing $k$ best chords while adhering to the connectivity requirement to creates the sequence of pins defining the chords.
In the following subsections, we formulate the space of chords, then define the optimization that calculates the fitness function of all possible chords, and then describe our method to select the chords for painting based on the fitness of chords.

\subsection{The space of chords}

\begin{figure}[tbp]
    \centering
    \includegraphics[height=2in]{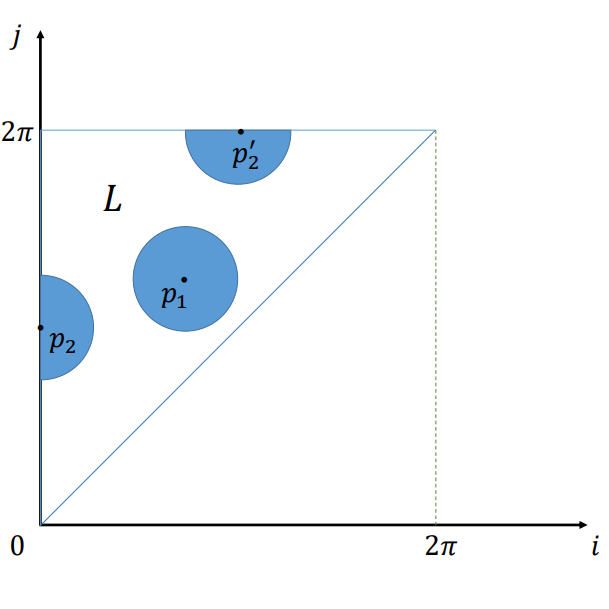}
    \caption{An illustration of the Chord space. The upper triangle is the valid parameter region. Two neighborhoods of points $p_1$ and $p_2$ are illustrated. Note that $p_2$ is equivalent to $p'_2$.}
    \label{fig:space}
\end{figure}

Let $L$ denote the set of all chords on a unit circle. Every element in $L$ can be represented by two parameters $(i,j)$, where $i,j \in \mathbb{R}$ represent the polar angle of the two endpoints. Since the direction of chords is not important, we can impose a requirement that $i \leq j$. Then, we can map a chord to an equivalence class $[(i,j)]$ formed by an equivalent relationship $\sim$ defined by the following rules:
\begin{itemize}
\item $(i,j)\sim(j,i)$
\item $(i,j)\sim(i+2\pi,j)$
\item $(i,j)\sim(i,j+2\pi)$
\end{itemize}

This means that we can concentrate on $i,j \in [0,2\pi)$, and the parameter space forms a right triangle $\lbrace (i,j) | 0 \leq i \leq j < 2\pi \rbrace$ which could be regarded as an identification topology deriving from the plane $\mathbb{R}^2$. Each chord in this space is represented by a point $(i,j)$. The distance between two chords $l_1=(i_1,j_1)$ and $l_2=(i_2,j_2)$  in this parameter space is defined as:
\begin{equation}
d(l_1,l_2)=\min_{(i,j) \in [i_1,j_1]} {\lbrace \max(|i-i_2|,|j-j_2|) \rbrace}
\end{equation}
which is the minimal $L_\infty$ distance between the two equivalence classes. An illustration of the chord space and the neighborhoods of chords is provided in Fig.\ref{fig:space}.

In our problem, the parameter space is discretized so that the endpoints can only be chosen in a finite set $\lbrace p_0, p_1,\dots,p_{P-1} \rbrace$ which represents the angles of the pin positions on the circumference of the circle. Thus, we obtain the corresponding discrete chord set $\tilde{L}=\lbrace (i,j) | i,j \in \mathbb{N}, 0 \leq i \leq j < P \rbrace$.

\subsection{Fitness computation}

With the formulated chord set $\tilde{L}$, our goal is to sample $k$ elements from this set in order, with the requirement that the end point of one chord is the start point of the next one. We seek a function $F:\tilde{L} \to \mathbb{R}$ that represents the fitness value for all of the possible chords. Suppose that there are $m$ chords in the set $\tilde{L} = \left\lbrace l_1, l_1, \dots, l_m \right\rbrace$. The fitness value $f_i$ for each chord $l_i$ is estimated by solving an optimization problem, and then $k$ chords will be sampled with priority for higher fitness values.

The fundamental thought is to reproduce the input image by assigning appropriate gray levels to each chord. Although our thread painting algorithm outputs two-value image, namely each pixel is either $0$ or $255$, at this stage, we loosen this restriction so that each chord can be assigned a real value defining its gray level, which will also serve as its fitness value.

The value of a reconstructed pixel is defined as the sum of gray-values of all chords covering it. For each pixel $I_i$ in the reversed circular image (we use a similar method of reversing black and white of the original image), we have:
\begin{equation}
I_i = f_{i_1}+f_{i_2}+\cdots +f_{i_s}
\end{equation}
where $f_{i_1},f_{i_2},\ldots,f_{i_s}$ are the gray-values of the chords $l_{i_1}, l_{i_2}, \dots , l_{i_s}$ that cover pixel $I_i$.

We can now define an overdetermined set of equations
\begin{equation}
Af=b
\label{simple}
\end{equation}
where $b=(I_1, I_2, \dots , I_n)$ is the vector of original (reversed) pixel values, $f=(f_1,f_2,\dots,f_m)$ are the fitness values to be determined, and $A$ is an $m \times n$ matrix where $A(i,j)=1$ if the $j$th chord covers the $i$th pixel, otherwise $A(i,j)=0$.

In some cases, the contrast of the input images are far from satisfactory, and sometimes users want to focus on the outlines and edges of the object in the picture. We modify the definition of the vector $b$ to accommodate this:
\begin{equation}
b_i = (1-\alpha) I_i + \alpha |g_i|
\label{equ_edge}
\end{equation}
where $g_i$ is the gradient vector at pixel $I_i$ and the parameter $\alpha$ controls the balance between gray level and gradient magnitude of the pixel that define the edge constraint. By default, we set $\alpha = 0$ in our experiment. When $\alpha>0$, the algorithm enhances the edges of the original image.

This formulation provides the same importance to every pixel. This means pixels with the same gray level and gradient magnitude will have similar importance, no matter where they appear in the original image. To allow for content-based importance, we extend our method by assigning different weights to each equation. Larger weights that are given to important regions in the image, will force the solver to provide more accurate reconstruction on the selected pixels, while sacrificing the accuracy of others. We define $W={\rm diag}(w_1,w_2,\dots,w_n)$ as the weight matrix, where each $w_i$ denotes the weight of pixel $i$ in the image. The objective function now becomes
\begin{equation}
E(f) = ||W \cdot (Af-b)||^2
\end{equation}
In our implementation, we use
\begin{equation}
w_i=\left\{
\begin{aligned}
2.0, & {\rm \ on \ important \ regions} \\
1.0, & {\rm \ otherwise}
\end{aligned}
\right.
\label{eq:importance}
\end{equation}

Minimizing the  objective function to get the fitness vector with only per-pixel constraints can give unstable results. Therefore, it is necessary to add regularization. We can achieve better results by giving a different weight to each chord. Let $V$ be a diagonal matrix ${\rm diag}(v_1,v_2,\dots,v_m)$, we define two terms related to chords:
\begin{equation}
v_i = \beta \exp(-s_i/P) + \gamma d_i / \max(d_i)
\end{equation}
The first term relates to the length of the chord and the second to the consistency between the chord and the edges in the original image.
$\beta$ and $\gamma$ are weighting parameters.

As short chords cover fewer pixels, they are less restricted in matrix $A$. Thus, their coefficients tend to be larger, resulting in excessive drawing on the marginal area of the circle, which is undesirable. We suppress the short chords by using the first term that depends on
$s_i$, the length of the chord $l_i$. If $l_i=\overline{p_{i_1}p_{i_2}}$, where $i_1,i_2$ are indices of pins, then $s_i$ is defined as the distance of two endpoints on the circle:
\begin{equation}
s_i = \min(|i_1-i_2|, P-|i_1-i_2|)
\end{equation}

The parameter $d_i$ reflects the consistency between the chord and the edges in the original image. We first calculate the unit direction vector $e_i$ for the chord $l_i$. Then, for each pixel $p$ covered by $l_i$, we compute the gradient vector $g_p$, and define an average consistency of chord $l_i$ as:
\begin{equation}
d_i = \frac{\sum_{p \in l_i} |g_p \times e_i|}{|l_i|}
\end{equation}
The cross product gives larger result if the chord is not aligned with the edges. Such chords are suppressed by our algorithm.

The complete formulation of the optimization problem is therefore defined as follows:
\begin{equation}
E(f) = ||W \cdot (Af-b)||^2 + ||V \cdot f||^2
\label{final}
\end{equation}
where $f=(f_1,f_2,\dots,f_m)$ are the unknowns representing the fitness values of all possible chords.

\subsection{Chord sampling}
\label{section:line_sampling}

With the fitness vector $f$ computed, a greedy sampling process can be defined. First, we set the number of chords $k$ by default to be:
\begin{equation}
k = 500 + 10 \cdot (255-{\rm avg}(I))
\end{equation}
where ${\rm avg}(I)$ is the average gray level value of the input image. This is intuitive since the darker the picture is, the more chords are  needed to paint it. However, this estimation is not always appropriate, so we allow the number of chords $k$ to be chosen by the user as well.

Next, given a starting pin $p_0$, the next pin $p_1$ in the sequence can be chosen as the one that maximizes the fitness of the chord $\overline{p_0p_1}$. This procedure can be repeated to collect all $k$ chords.

This greedy procedure tends to produce excessive chords over dark region. To relieve this accumulative effect, we leverage an error diffusion procedure. First, we map the fitness value into the range $[0,1]$ by using:
\begin{equation}
f_i \leftarrow \frac{1}{2}\left(\tanh(f_i/T) + 1\right)
\end{equation}
Note that the raw coefficient for each chord can also be negative, as this is not prohibited in the optimization step. Thus, a normalization step is necessary. The error created by $l_i$ is $1-f_i$. This error value will be evenly diffused to the $\epsilon$-neighbourhood of $l_i$ : $N_{\epsilon}(l_i)=\lbrace l_j | d(l_i,l_j) \leq \epsilon \rbrace$, i.e., the fitness value of each element in $N_{\epsilon}(l_i)$ is decreased by $(1-f_i)/|N_{\epsilon}(l_i)-1|$. The parameter $T$ controls the amount of error to be propagated. if $T$ is smaller, the chosen fitness value will be closer to $1$, so less error will be diffused, leading to more intensive result. If $T$ is larger, the value of the neighboring chords will be significantly reduced so the selected chords tend to be separately distributed. The value of $T$ is adjustable by the user.

%


\section{Experiment}

We collected 15 pictures for evaluation, including portraits, sketches and cartoons (see Table~\ref{tab:SSIM_comp}). To reduce the computational workload, we resize the input square region to be $401 \times 401$ (i.e. radius $200$ for the circle) and set the number of pins $P=300$. By default we fix $\alpha = 1.0$, $\beta = 5.0$, $\gamma = 10.0$, $T=30.0$, and $\epsilon=2$. Some examples of the results are displayed during the following discussion, and others are shown in Fig.\ref{fig:more_examples}.

\begin{table}[tbp]
\centering
        \resizebox{\textwidth}{!}{\begin{tabular}{l|c|c|c|c|c|c}
\hline
& \multicolumn{3}{c|}{SSIM (original)} & \multicolumn{3}{c}{SSIM (blurred)} \\
\cline{2-7}
\multirow{2}{*}{Image Name} & \multirow{2}{*}{Greedy} & \multirow{2}{*}{Ours} & Ours  & \multirow{2}{*}{Greedy} & \multirow{2}{*}{Ours} & Ours \\
& & & (disconnected) & & & (disconnected) \\
\hline
Jerry      & 0.216 & \textbf{0.266} & 0.261 & 0.376 & 0.383 & \textbf{0.384} \\
Winnie     & 0.205 & \textbf{0.258} & 0.256 & 0.434 & 0.388 & \textbf{0.405} \\ 
Girl 1     & 0.331 & \textbf{0.373} & 0.358 & 0.539 & \textbf{0.576} & 0.565 \\ 
Girl 2     & 0.311 & 0.348 & \textbf{0.352} & 0.506 & 0.522 & \textbf{0.542} \\ 
Poetin     & 0.281 & \textbf{0.308} & 0.302 & 0.458 & 0.473 & \textbf{0.478} \\
Trump      & 0.221 & \textbf{0.257} & 0.255 & 0.414 & 0.433 & \textbf{0.436} \\
Van Gogh   & 0.268 & \textbf{0.360} & 0.346 & 0.418 & \textbf{0.475} & 0.463 \\
Du Fu      & 0.265 & \textbf{0.310} & 0.303 & 0.433 & 0.445 & \textbf{0.451} \\
Leaf       & 0.251 & \textbf{0.313} & 0.302 & 0.447 & \textbf{0.481} & 0.468 \\
Flower     & 0.220 & \textbf{0.251} & 0.248 & \textbf{0.427} & 0.399 & 0.408 \\
Nuclear    & 0.250 & \textbf{0.274} & 0.260 & 0.339 & \textbf{0.345} & 0.330 \\
Leonardo   & 0.275 & \textbf{0.309} & 0.299 & 0.428 & \textbf{0.474} & 0.469 \\
Jobs       & 0.206 & \textbf{0.248} & 0.247 & 0.326 & \textbf{0.329} & 0.326 \\
Mario      & 0.229 & \textbf{0.274} & 0.267 & 0.374 & \textbf{0.384} & 0.384 \\
Mushroom   & 0.236 & \textbf{0.271} & 0.256 & 0.416 & \textbf{0.441} & 0.425 \\
\hline \\
        \end{tabular}}
\caption{SSIM index of different methods. See details in text.}
\label{tab:SSIM_comp}
\end{table}

\subsection{Similarity comparison}

We use SSIM~\cite{wang2004image} to quantitatively assess the quality of the chord representation. Although the resolution of our output image is $1001 \times 1001$, we resized both the input and output to be $201 \times 201$. To reduce the difference between the gray level input and the binary-valued output, we conducted an additional test on a blurred version of all samples. The number of chords to be drawn by all of the compared methods were the same. However, the greedy method (proposed in~\cite{ttweb}) sometimes terminated earlier because the sum of pixel values on the next chord became negative. Moreover, in this test all of the pixels in the image had the same weights (i.e., we set $W$ to be an identity matrix in Eq.\eqref{final}).

\begin{figure}[tbp]
    \centering
    \begin{tabular}{c@{\hspace{1mm}}c@{\hspace{1mm}}c@{\hspace{1mm}}c@{\hspace{1mm}}c@{\hspace{1mm}}c}
            \includegraphics[height=1.1in]{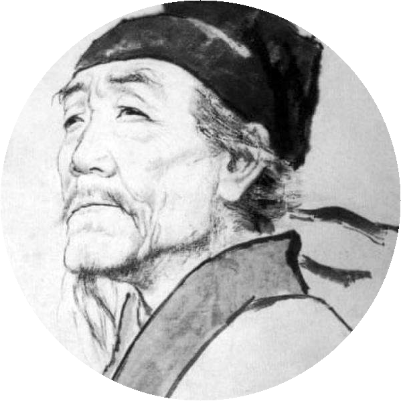}&
            \includegraphics[height=1.1in]{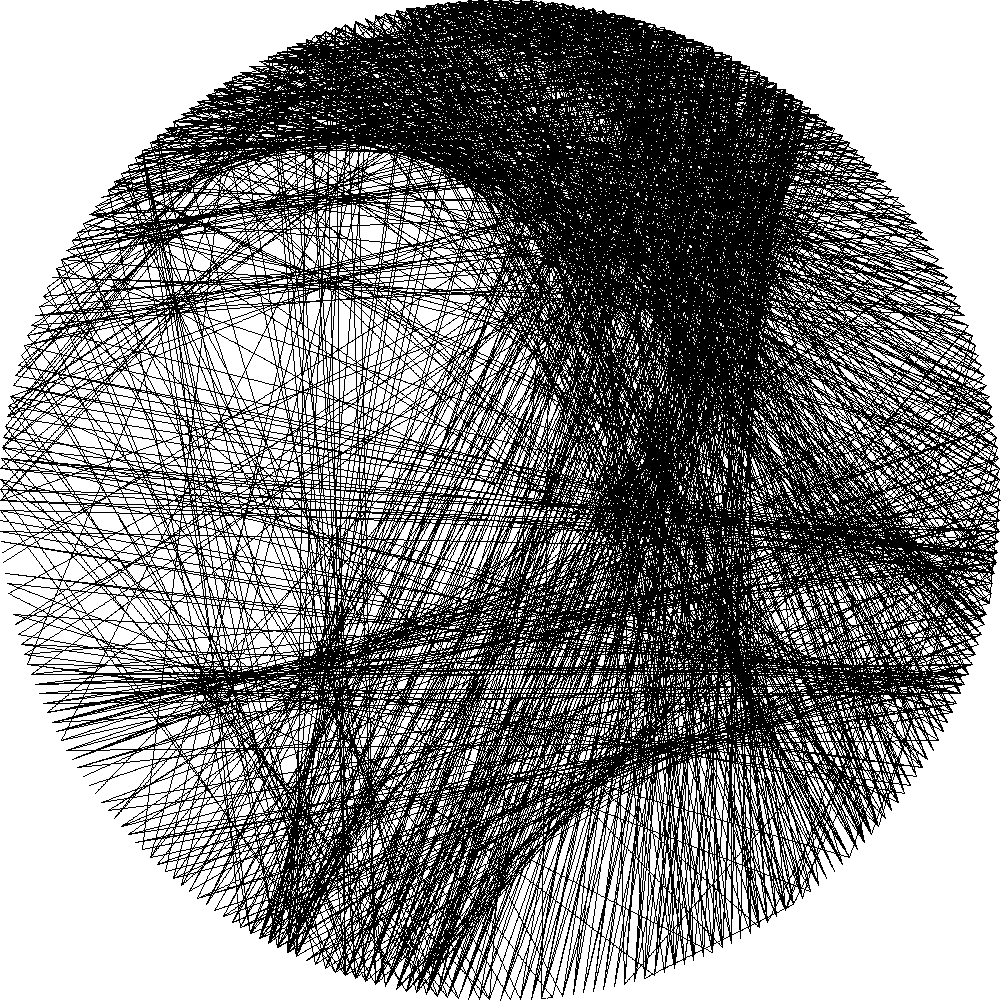}&
            \includegraphics[height=1.1in]{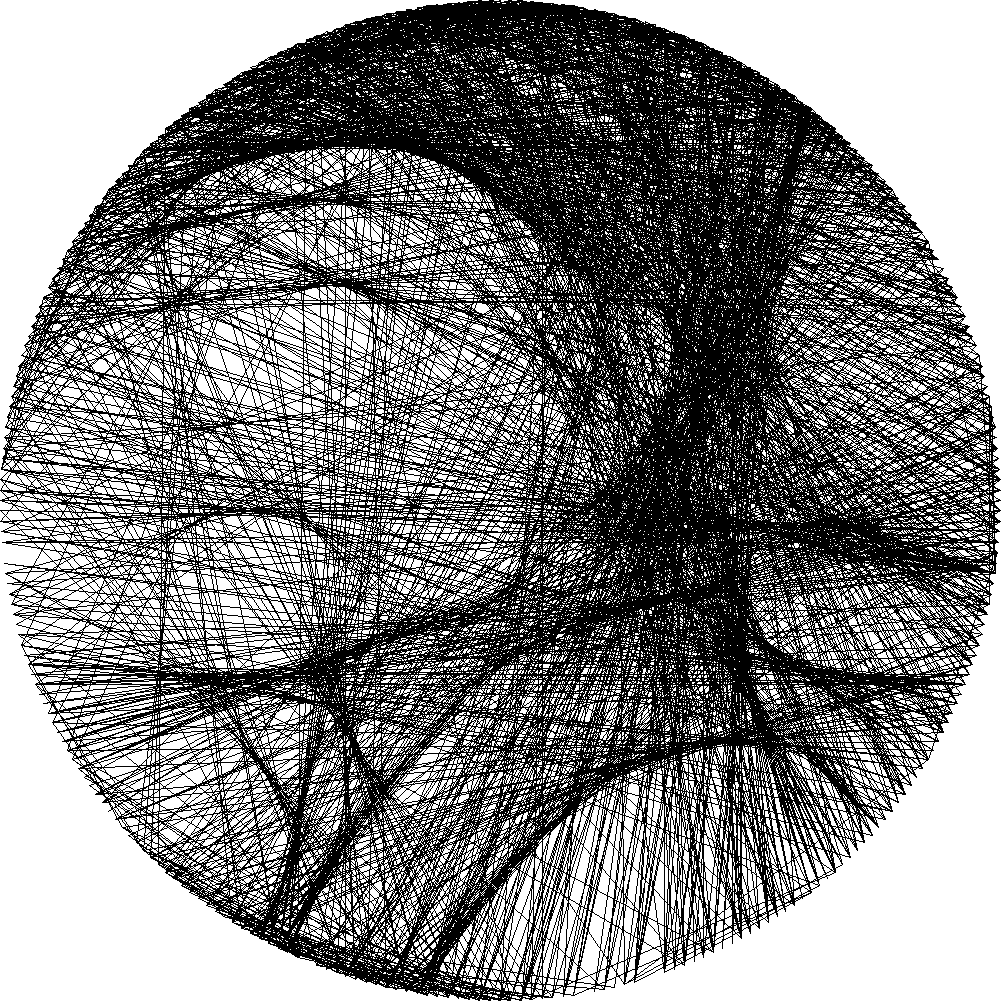}&
            \includegraphics[height=1.1in]{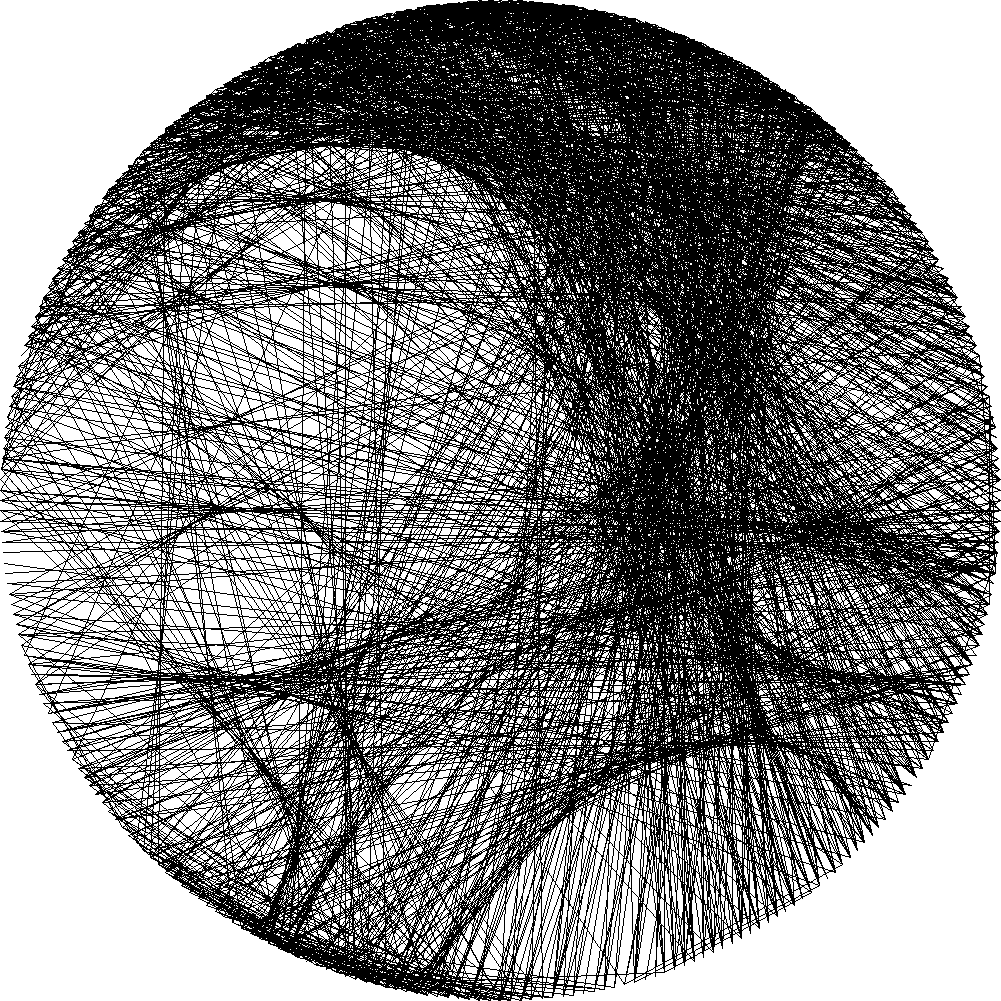}\\

            \includegraphics[height=1.1in]{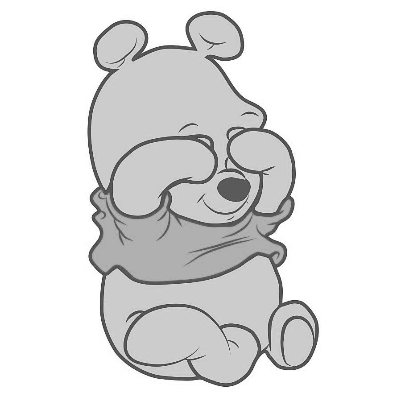}&
            \includegraphics[height=1.1in]{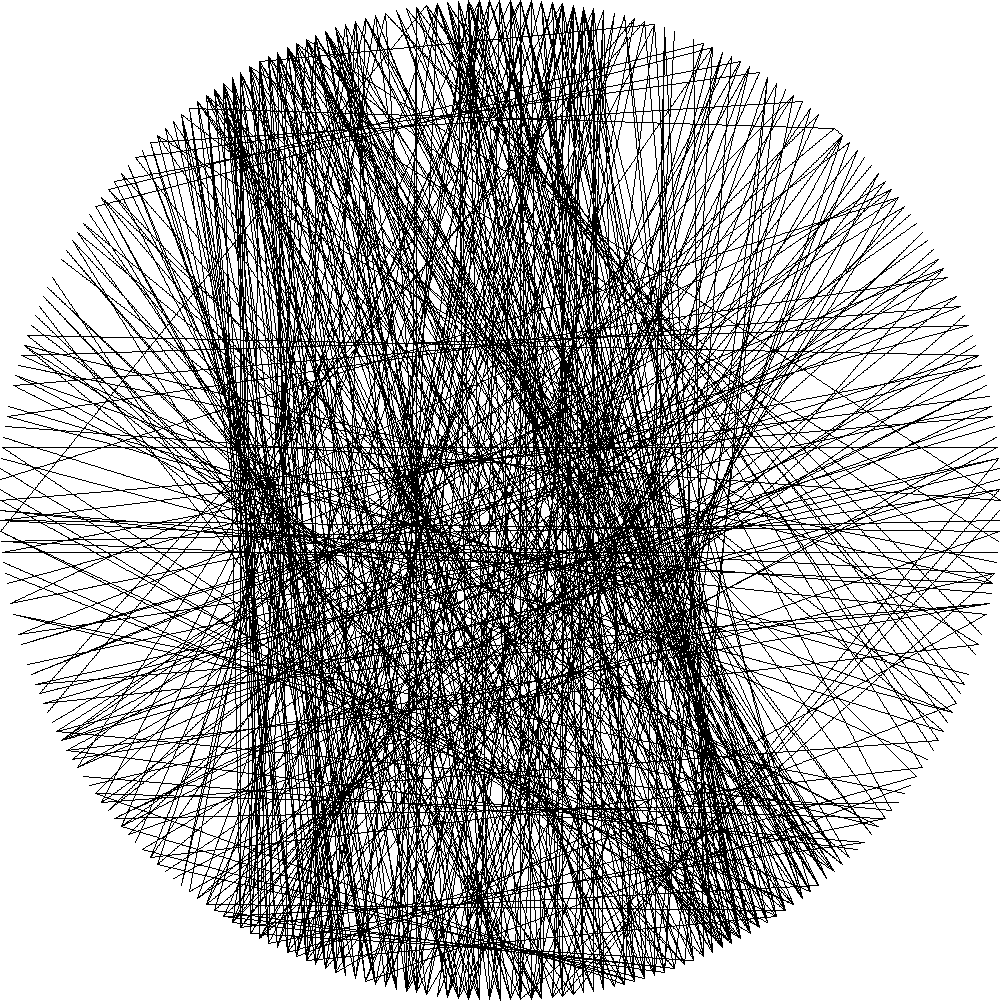}&
            \includegraphics[height=1.1in]{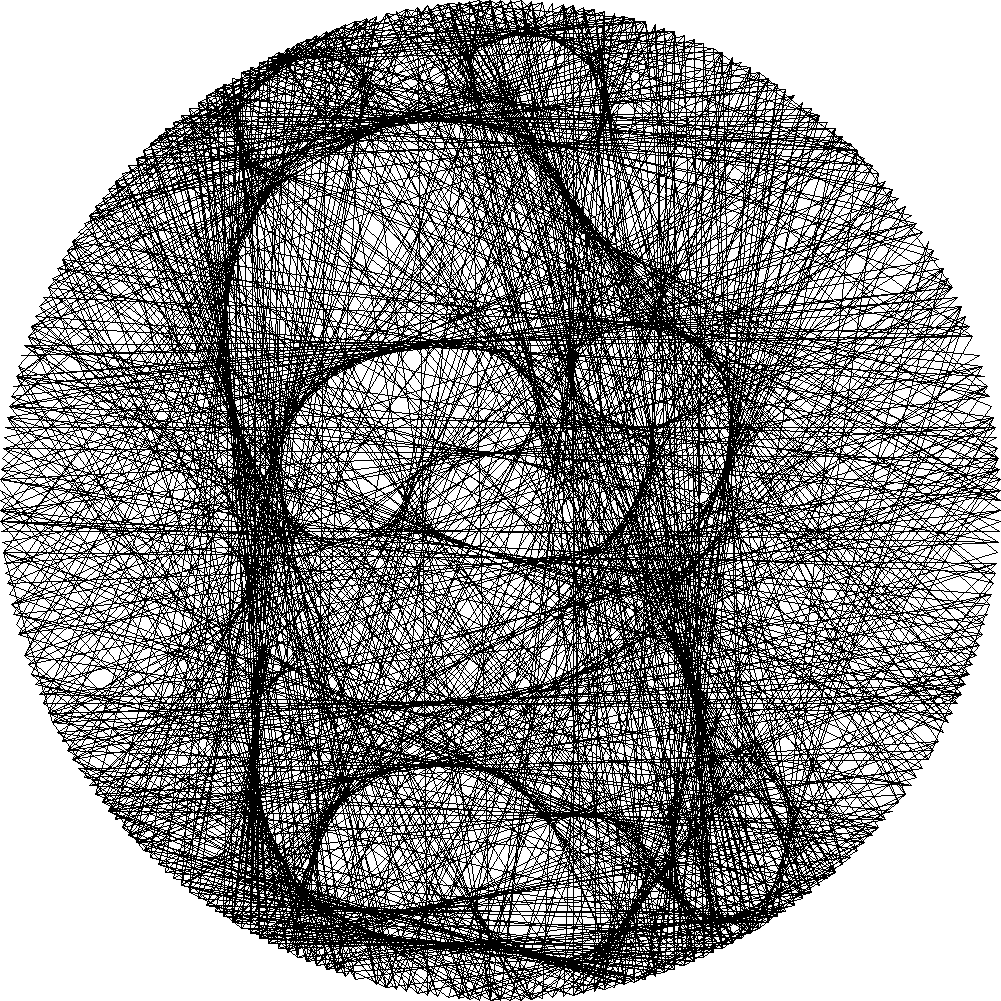}&
            \includegraphics[height=1.1in]{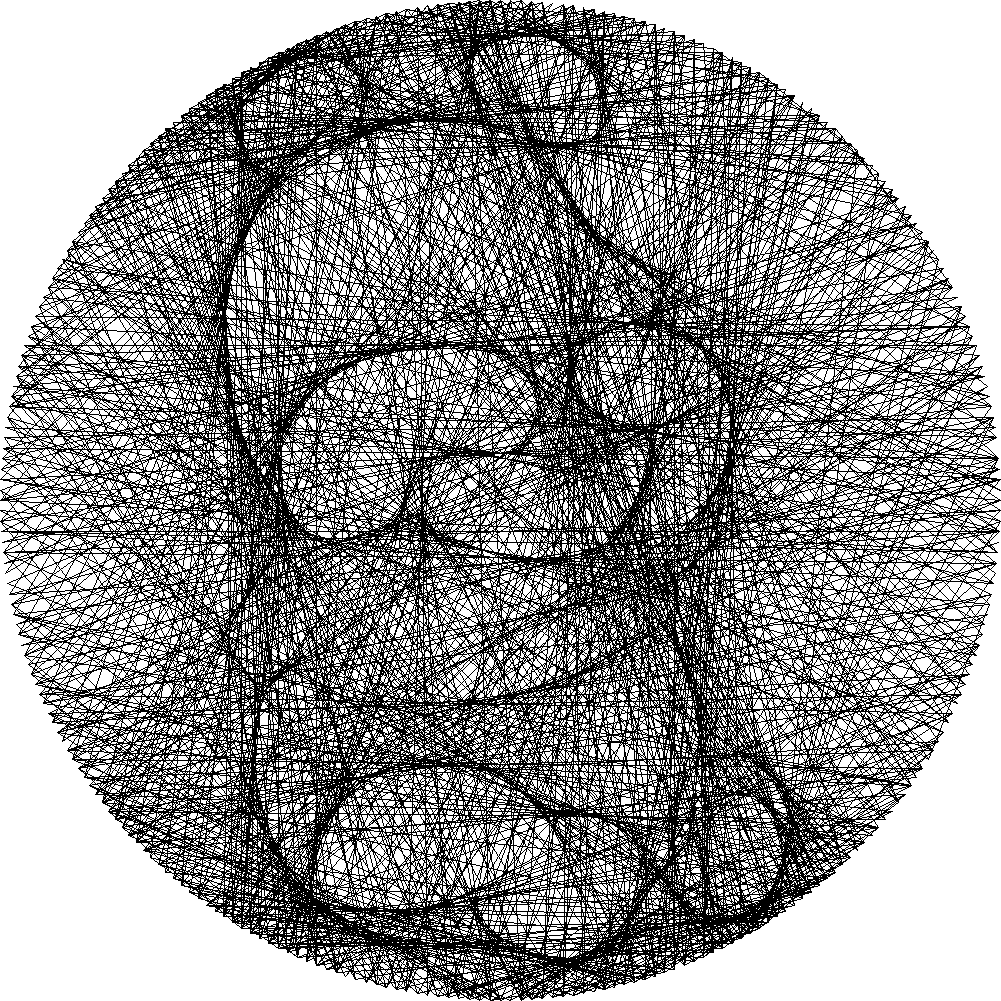}\\

            \includegraphics[height=1.1in]{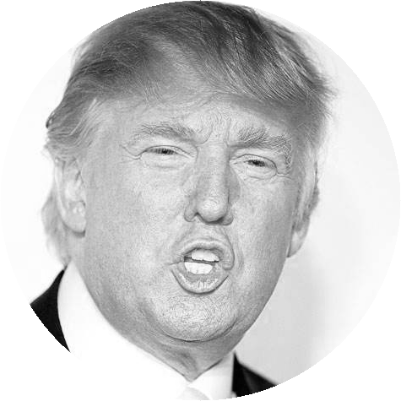}&
            \includegraphics[height=1.1in]{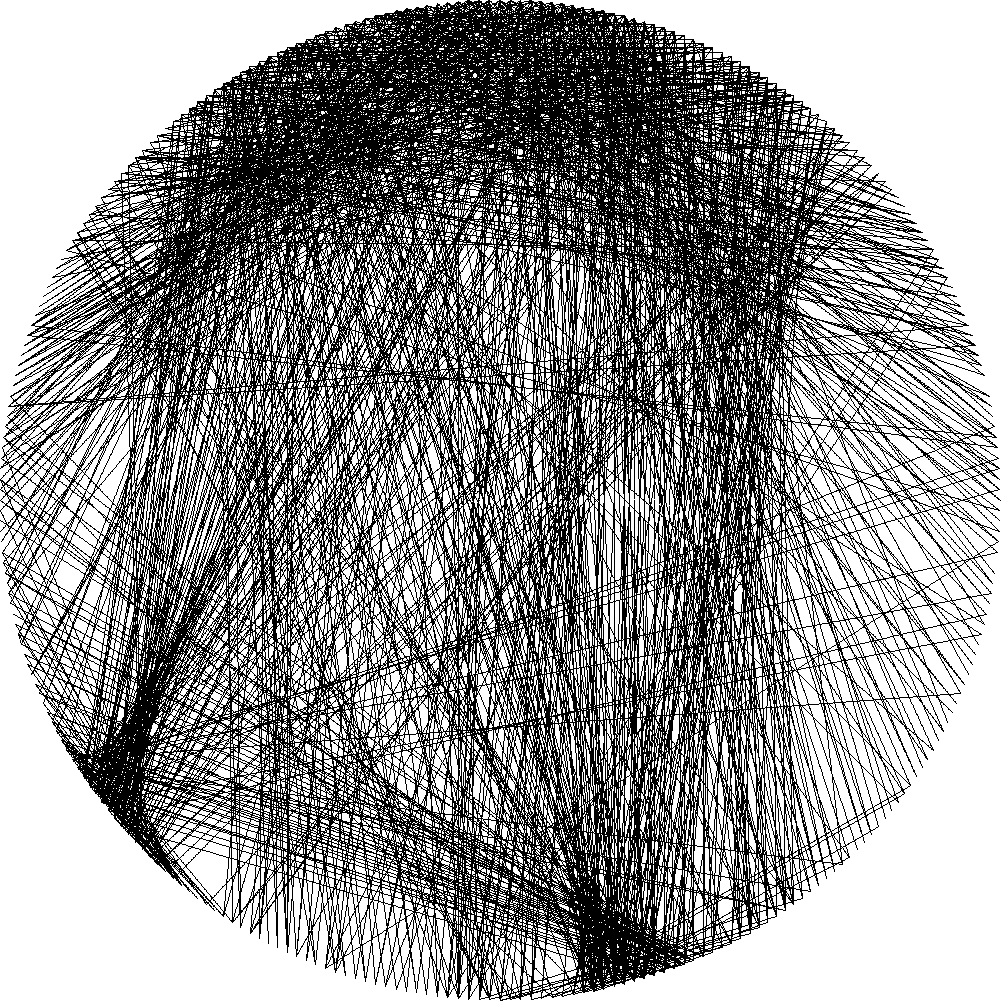}&
            \includegraphics[height=1.1in]{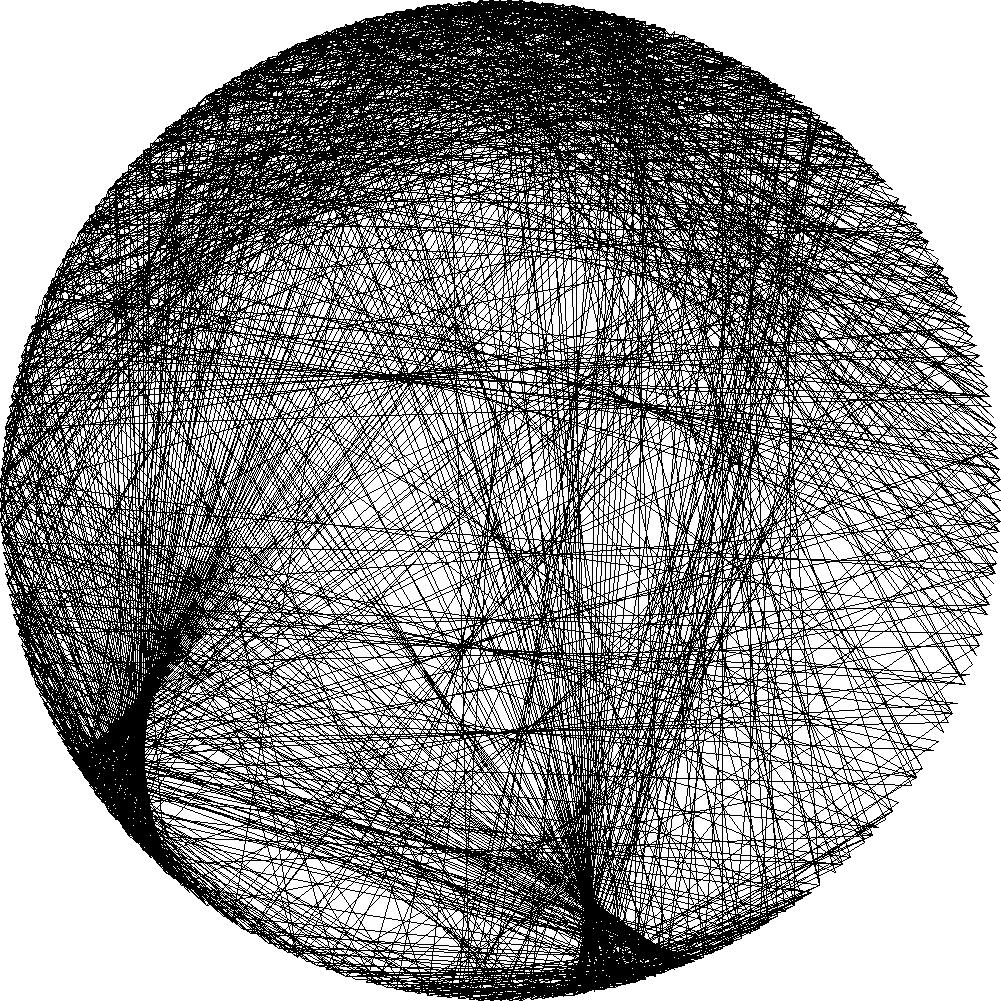}&
            \includegraphics[height=1.1in]{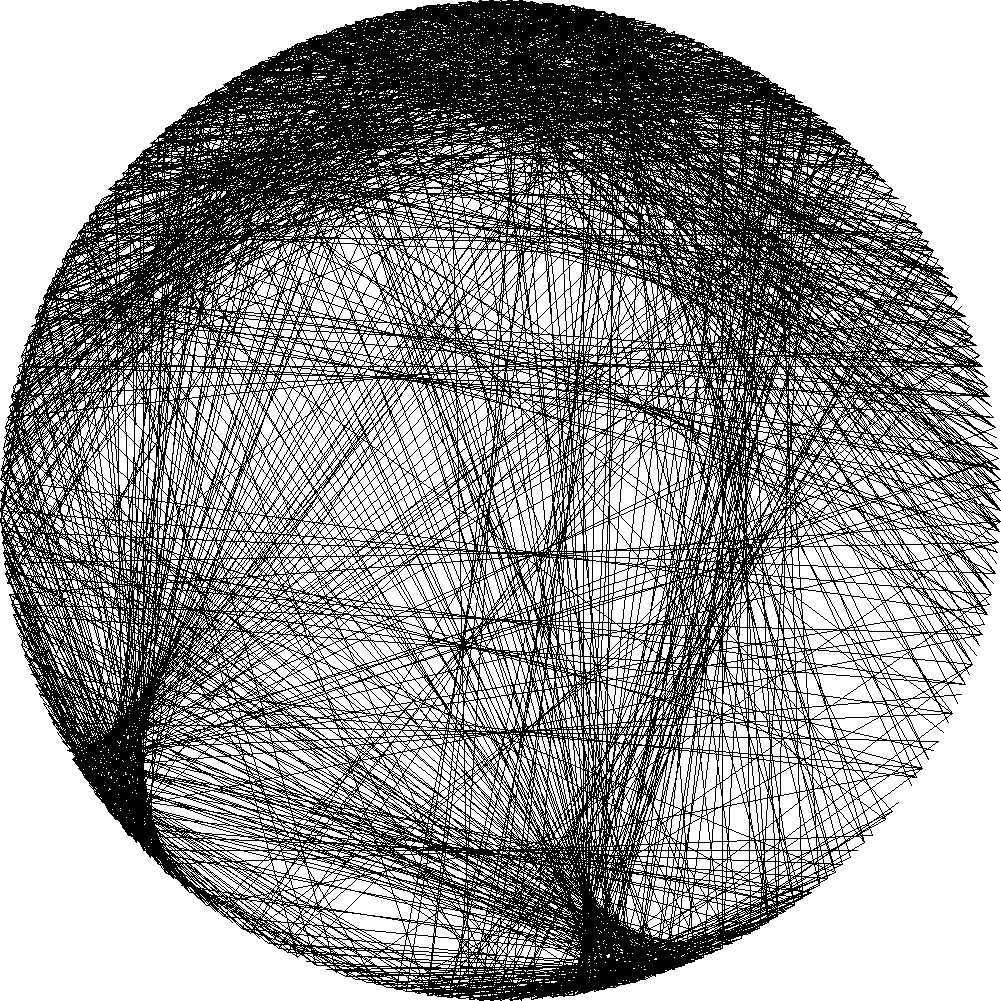}\\

            \includegraphics[height=1.1in]{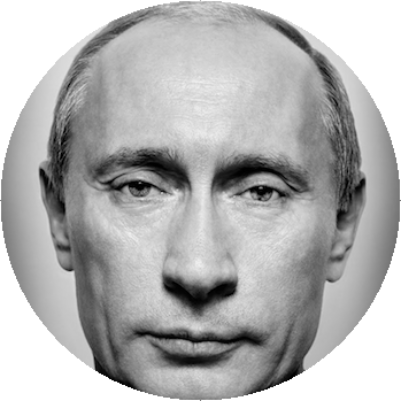}&
            \includegraphics[height=1.1in]{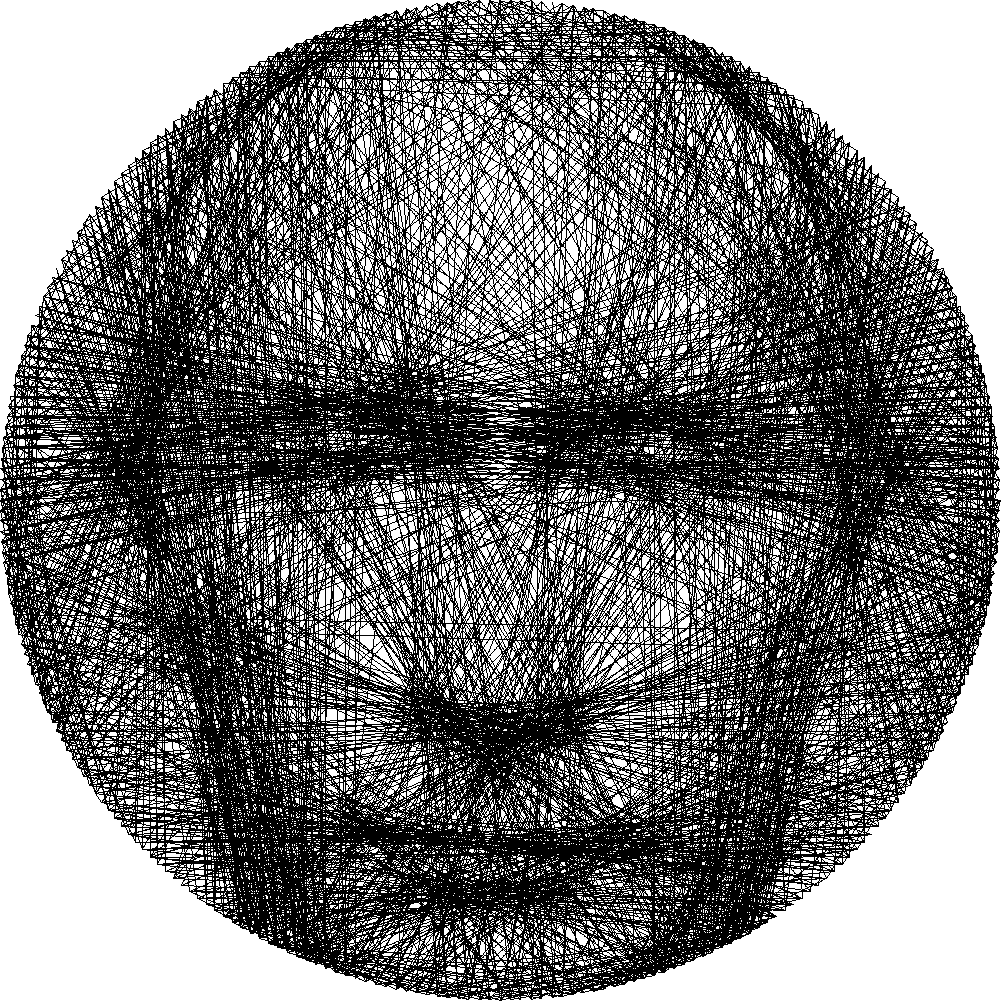}&
            \includegraphics[height=1.1in]{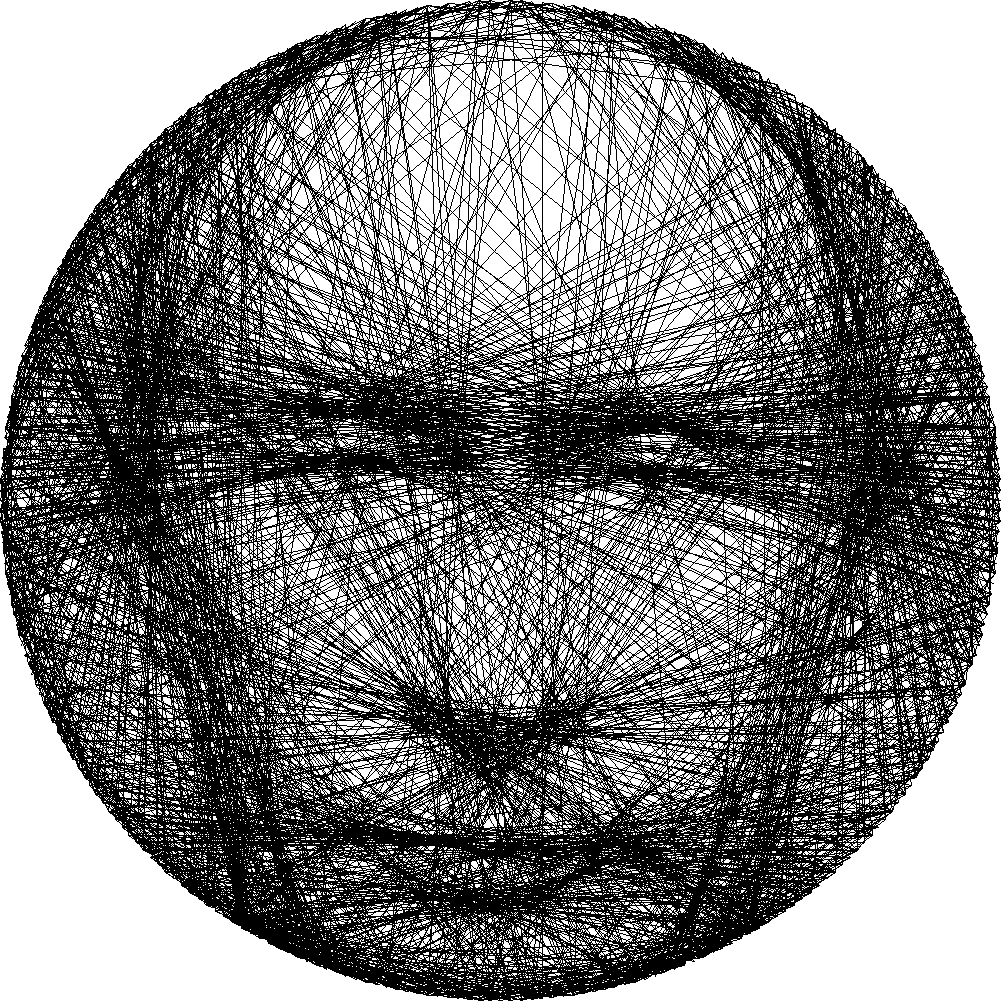}&
            \includegraphics[height=1.1in]{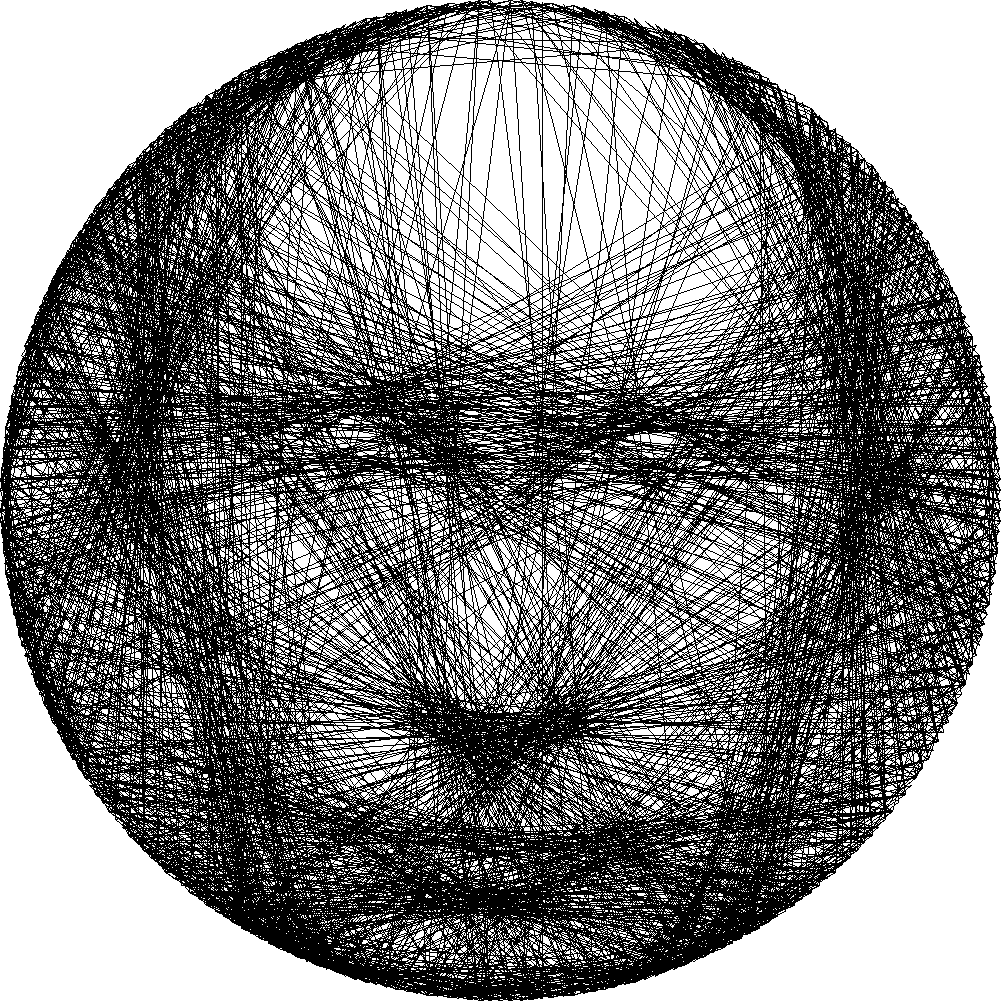}\\
            (a) & (b) & (c) & (d) \\

    \end{tabular}
    \caption{More examples of thread painting (Row 1-4: Du Fu, Winnie, Trump, Poetin). (a) Original image; (b) Results of greedy method; (c) Results of our method; (d) Results of our method (without connectivity requirement). }
    \label{fig:more_examples}
\end{figure}

Table~\ref{tab:SSIM_comp} compares the SSIM index of the outputs of different sampling strategies. We illustrate the result of 4 examples in Fig.\ref{fig:more_examples}. Our sampling method achieved higher similarity than the greedy sampling method.
We also tested another sampling strategy that loosens the connectivity requirement. If we do not require the chords to be connected in sequence, we can use a priority queue to choose the best chords. In each iteration, we select the chord with highest fitness value and remove it from the queue. Then, we update the fitness vector by decreasing the value of its neighboring chords according to the diffused error. These two sampling strategies (connected vs disconnected) are also compared in Table~\ref{tab:SSIM_comp}. Interestingly, removing the connectivity requirement actually reduces the SSIM value most of the time (for the test without blurring). This is because in this setting, the algorithm tended to select shorter chords that have larger fitness value. Therefore, the central area is inadequately painted. In this setting the user must increase the number of drawn chords to complete the thread painting. In the blurred comparison, all of the similarity values increase and the gap between the different methods is narrowed.

It should be noted that a thread painting is an abstract representation of the original image with dramatic differences. Hence, the traditional similarity measuring indices such as SSIM and PSNR for these images are relatively low. Therefore, we conducted a user study to judge the quality of the resulting images, which is described in Section~\ref{section:userstudy}.

\begin{figure}[tbp]
    \begin{tabular}{cc}
        &
        \begin{tabular}{ccccc}
            \includegraphics[height=0.8in]{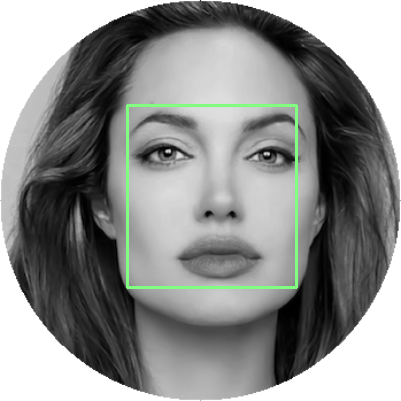}&
            \includegraphics[height=0.8in]{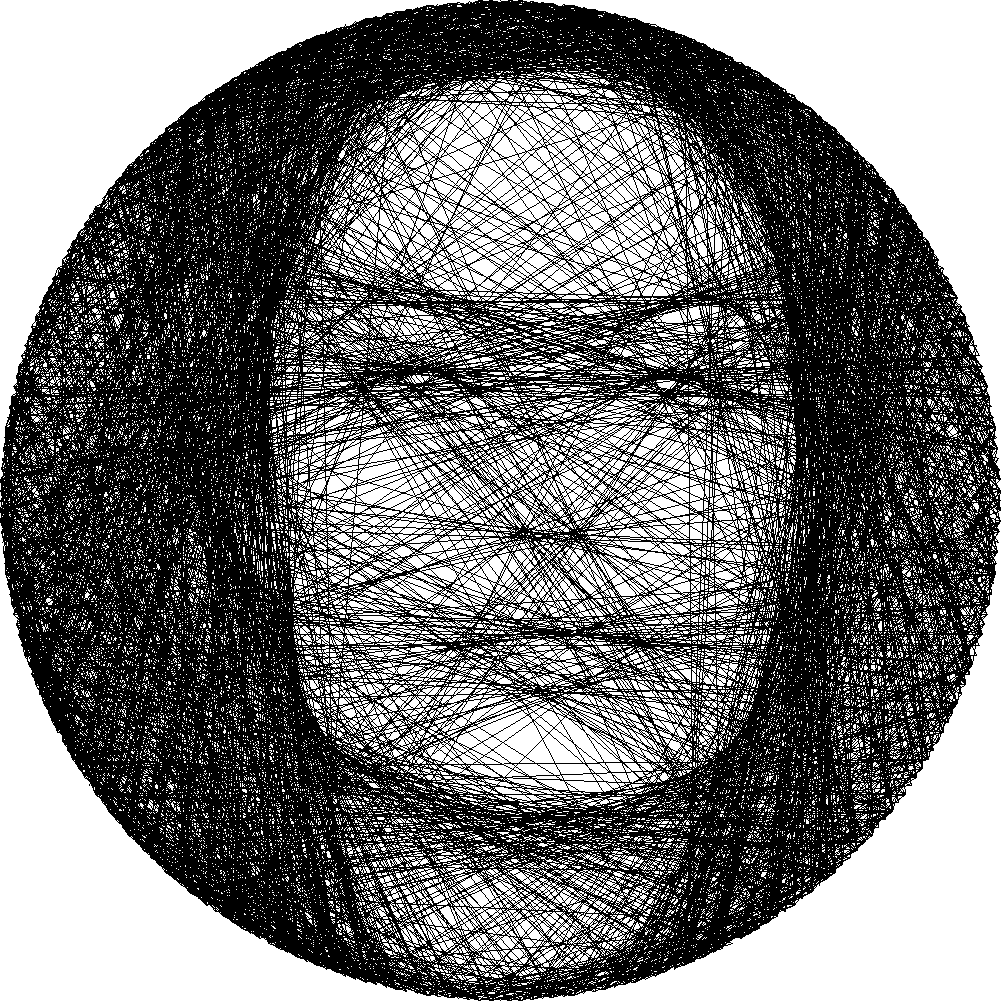}&
            \includegraphics[height=0.8in]{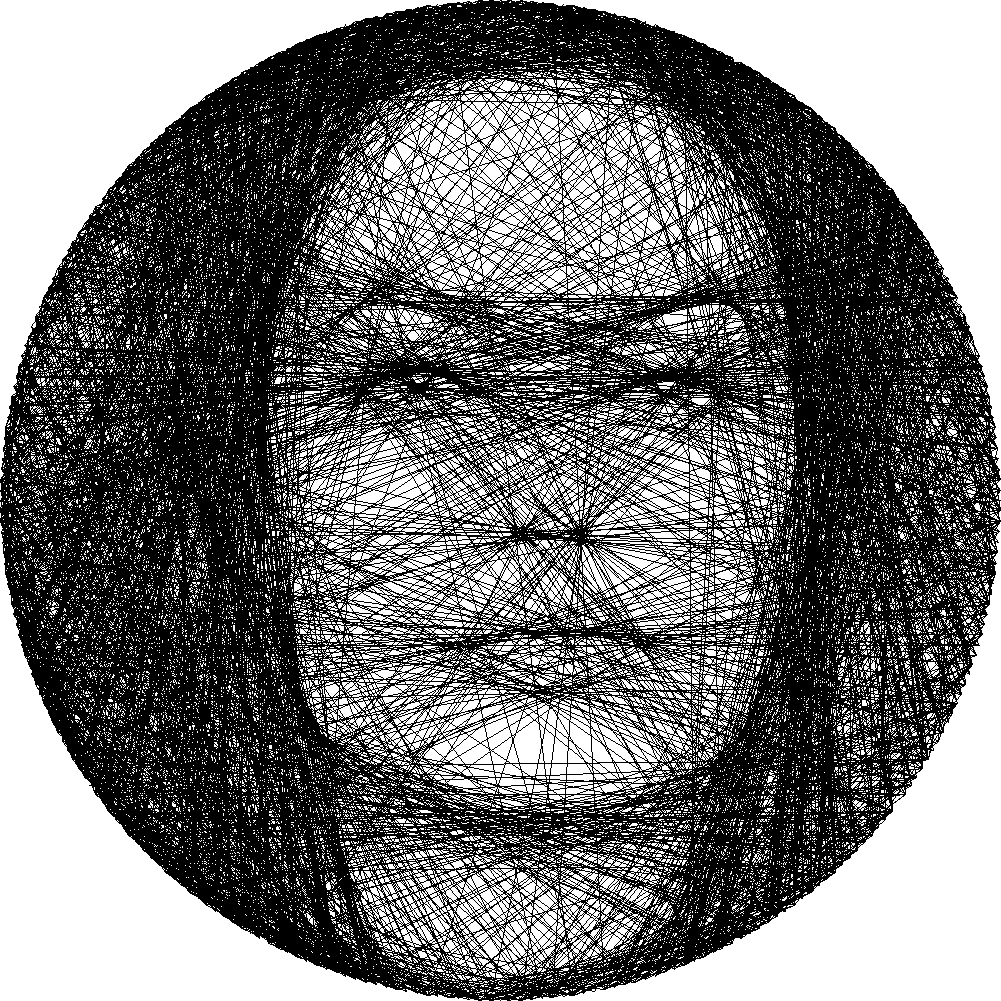}&
            \includegraphics[height=0.8in]{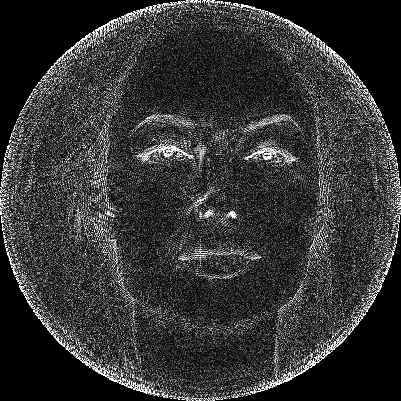}&
            \includegraphics[height=0.8in]{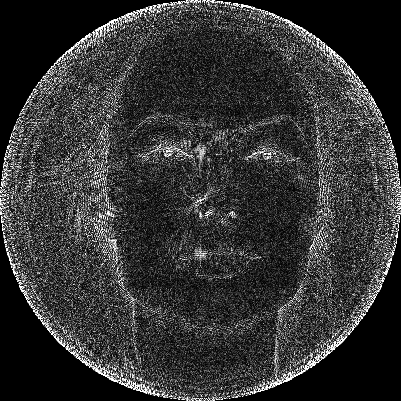}\\
        \end{tabular}\\
        &
        \begin{tabular}{ccccc}
            \includegraphics[height=0.8in]{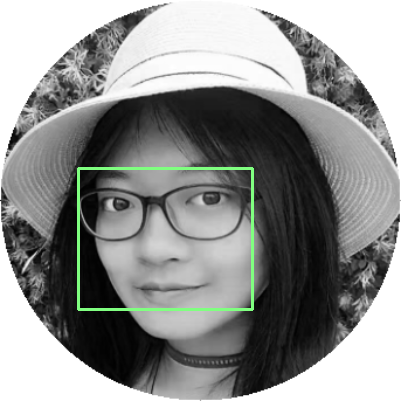}&
            \includegraphics[height=0.8in]{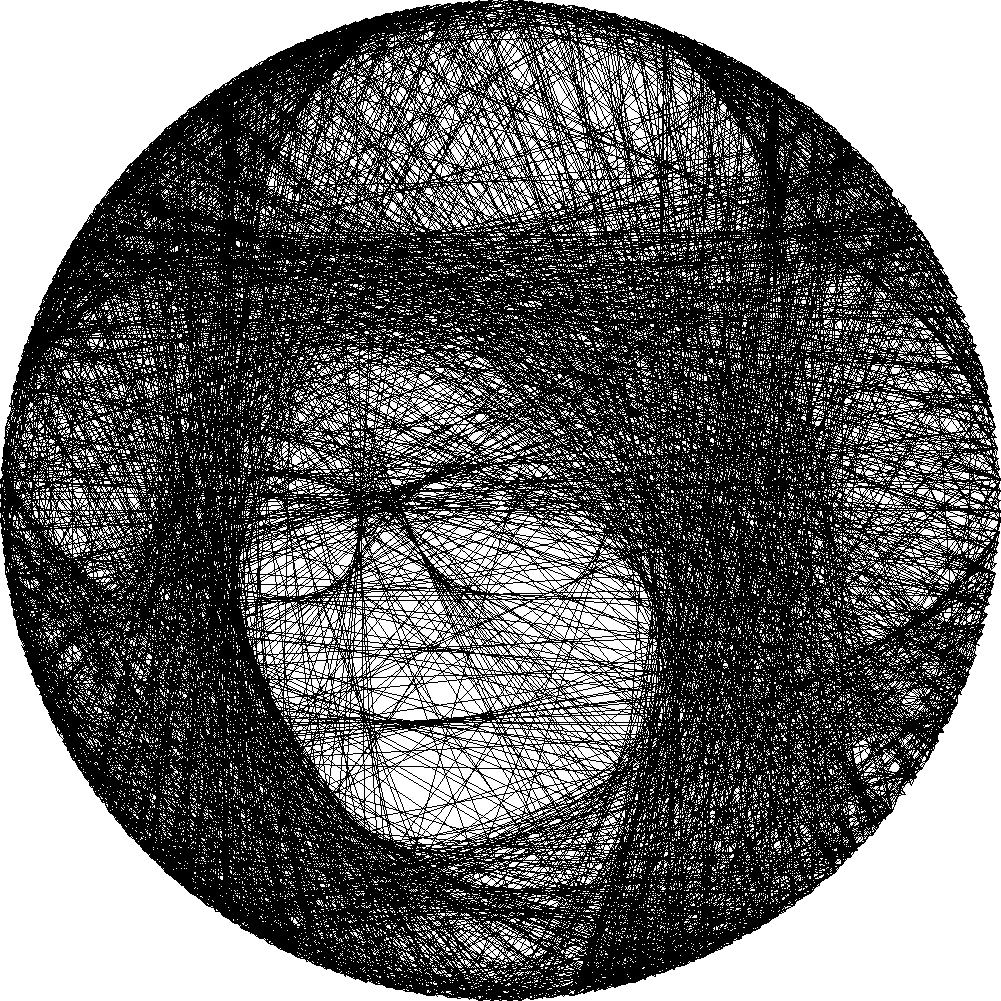}&
            \includegraphics[height=0.8in]{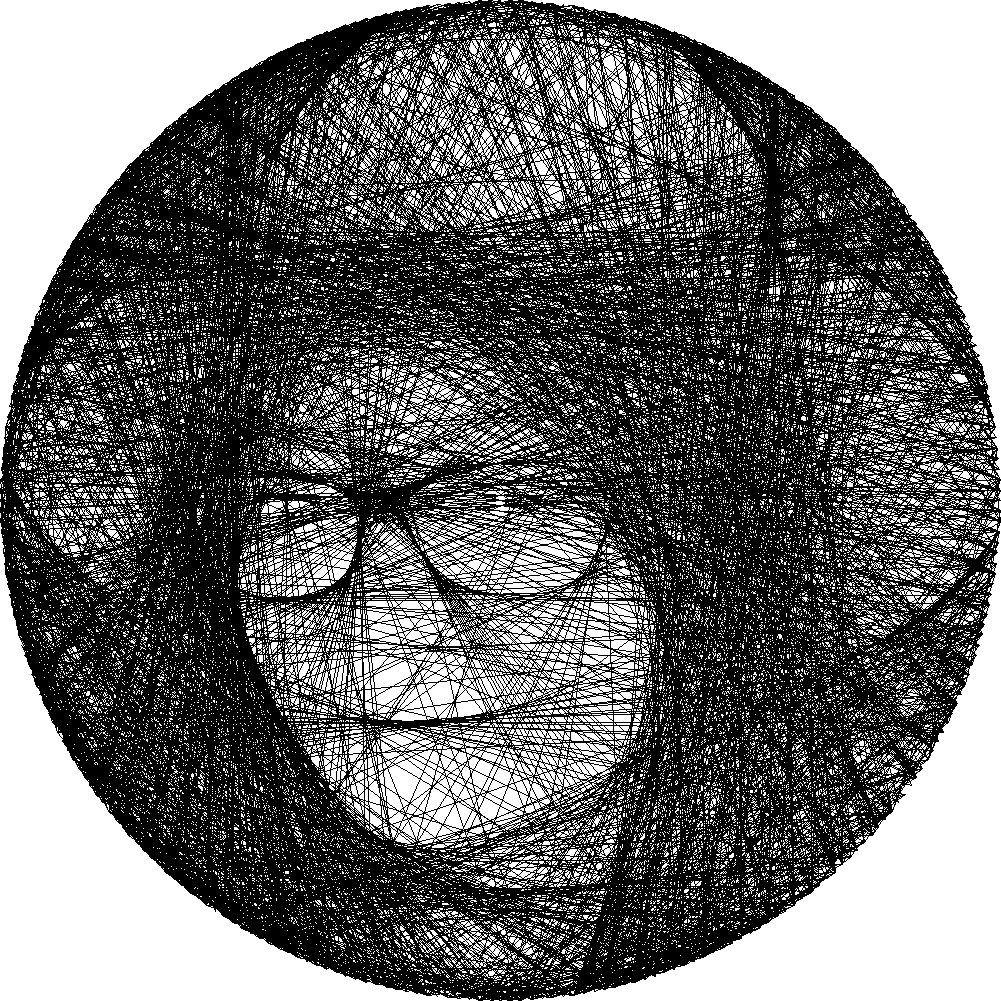}&
            \includegraphics[height=0.8in]{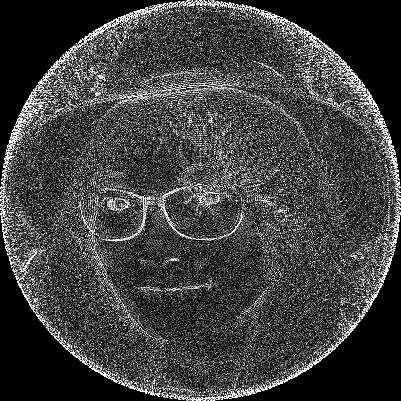}&
            \includegraphics[height=0.8in]{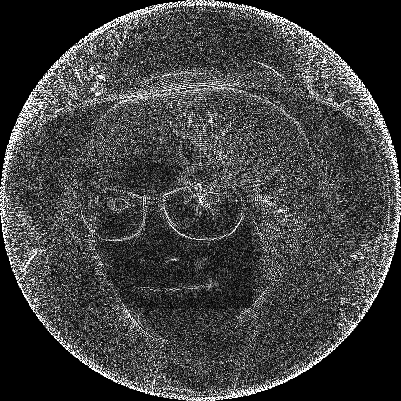}\\
            (a) & (b) & (c) & (d) & (e) \\
        &
        \end{tabular}\\
    \end{tabular}
    \caption{Example of the effect of adding weights on important region (row 1: Girl 1, row 2: Girl 2). (a) the region marked with the green rectangle is emphasized as important; (b) result with uniform weights; (c) result with importance based weights (Eq.~\ref{eq:importance}); (d) error map with uniform weights; (e) error map with importance weights. The error maps are computed between the input and the reconstruction of the whole chord set $\tilde{L}$. The error in the maps is amplified by a factor of 5 for visibility.}
    \label{fig:example_weights}
\end{figure}

\begin{figure}[tbp]
    \centering
    \begin{tabular}{c@{\hspace{1mm}}c@{\hspace{1mm}}c@{\hspace{1mm}}c@{\hspace{1mm}}c@{\hspace{1mm}}c}
        \includegraphics[height=1.1in]{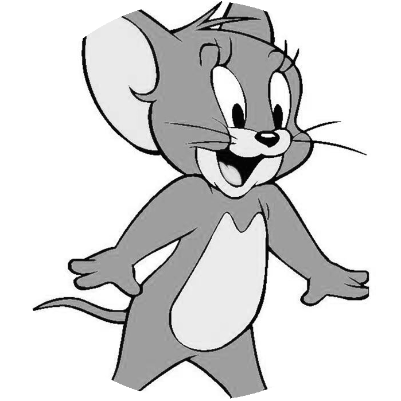}&
        \includegraphics[height=1.1in]{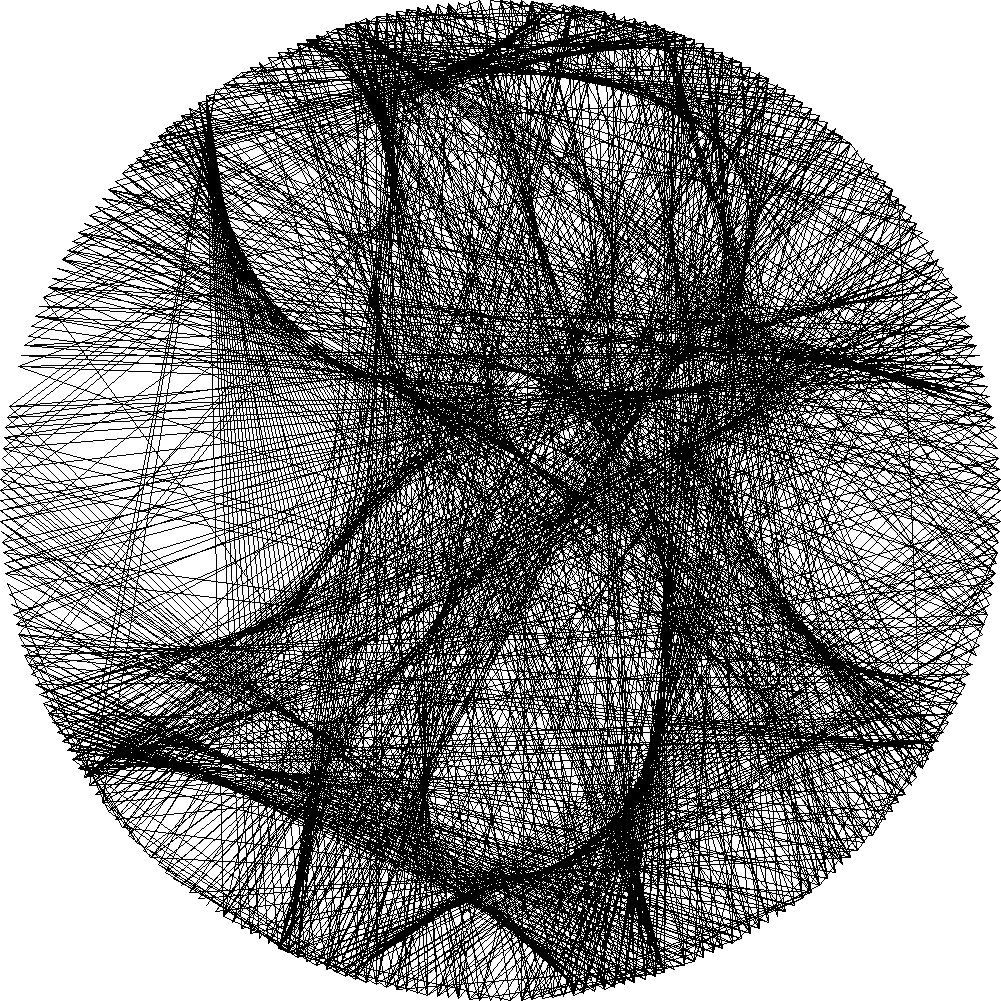}&
        \includegraphics[height=1.1in]{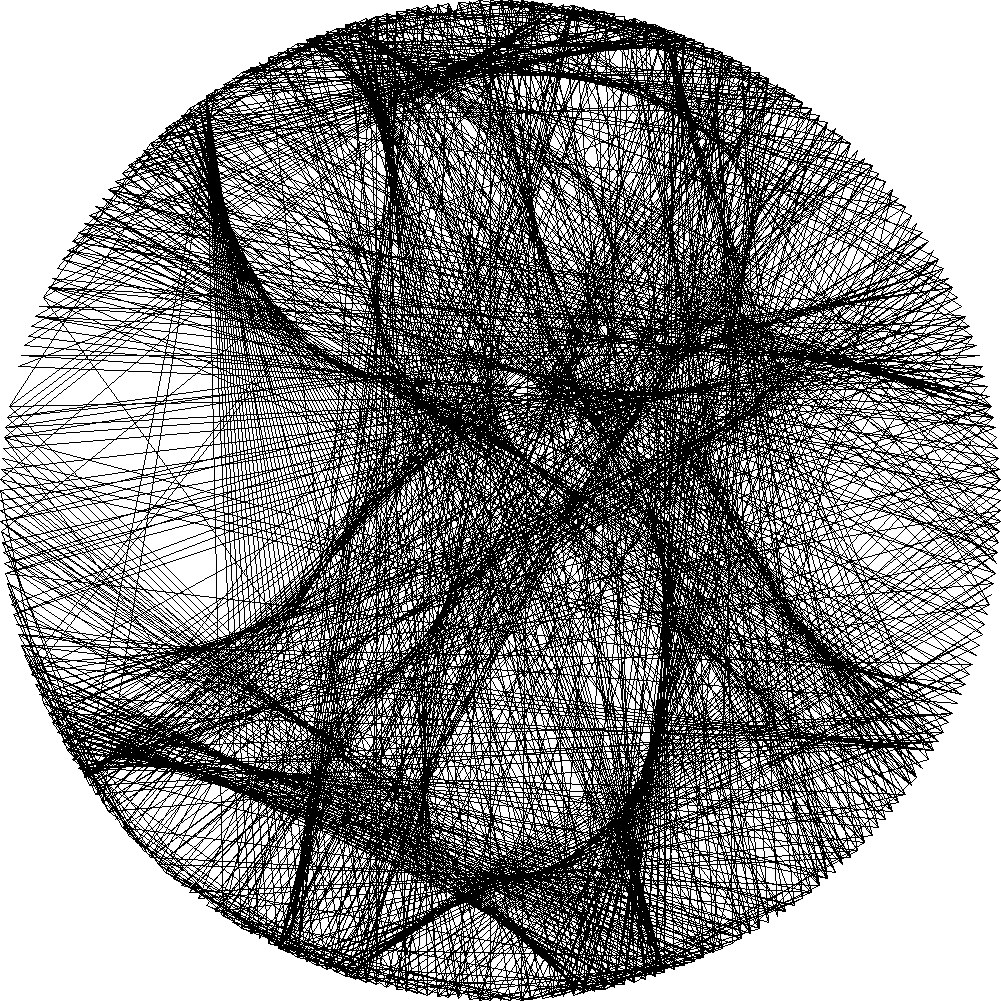}&
        \includegraphics[height=1.1in]{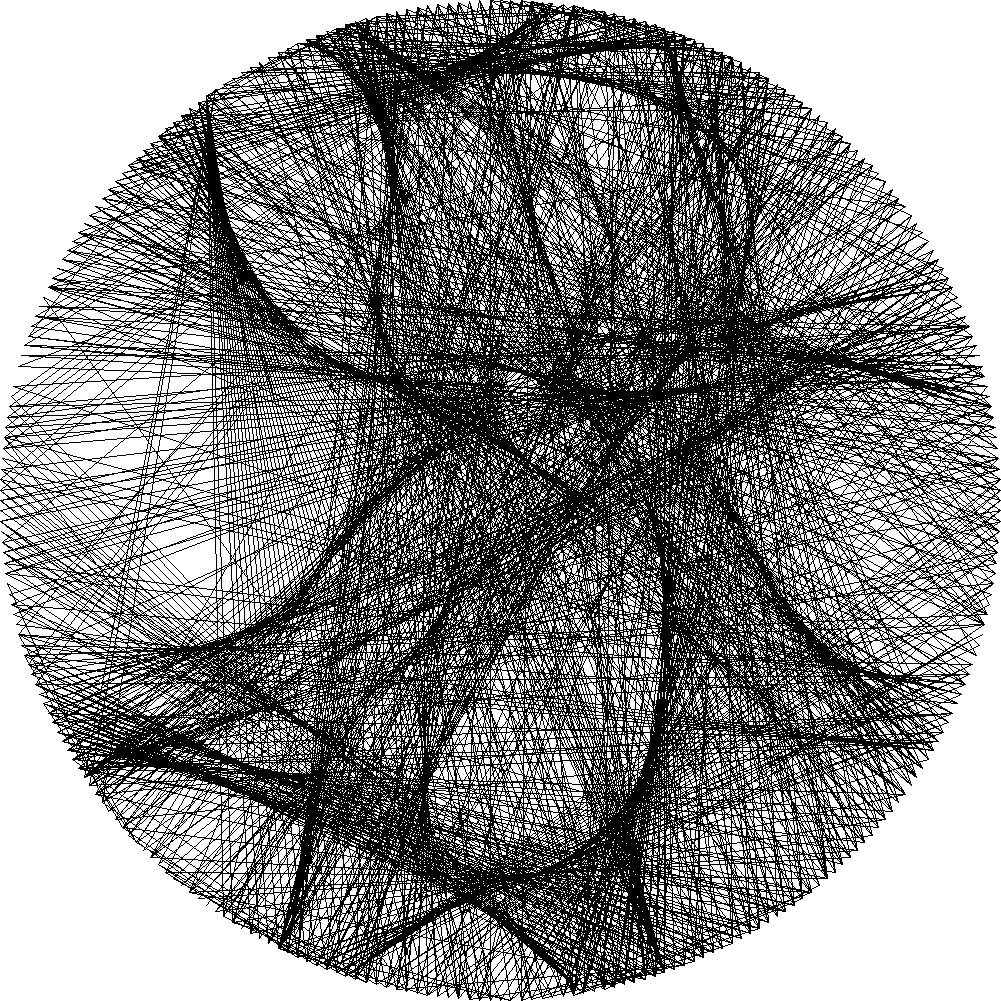}\\
        (a) & (b) & (c) & (d) \\
    \end{tabular}
    \caption{Illustration of the effect of the regularization term on the Jerry example. (a) original image; (b) result without the regularization term measuring direction; (c) result with identical weight on all chords ($V$ is the identity); (d) result with all regularization terms.}
    \label{fig:regular_weight}
\end{figure}

\subsection{Effect of user specified weights}

In many circumstances, the user may put more emphasis on a certain part of the image.  The quality of approximation on this part becomes more important than other parts. For example, the face region in a portrait is more important than the background. Using different weights on pixels of different regions (Eq.\eqref{eq:importance}) allows the optimization to put more emphasis on these regions. Fig.\ref{fig:example_weights} provides some results with weighted input. The eyes and glasses of the people were enhanced after marking the important region in the face.

We computed the error map measuring the difference between input image and the reconstruction from the fitness vector of chords. As shown in Fig.\ref{fig:example_weights}, the error in the marked region is reduced, while the error outside this region is increased. Transferring error from salient region to unimportant region can improves the perceived visual quality.

\subsection{Effect of regularization}

We conducted experiments on the effect the regularization terms. Fig.~\ref{fig:regular_weight} displays an example. When we removed the term that measures the fitness of the chord direction, the edges in the original image are preserved less. If the matrix of weights of chords is set to be the identity, the optimization solver tends to give high fitness value to the short chords, resulting in excessive drawing near the circumference of the circle. The short chords have less restrictions in the linear system, so it is essential to apply stronger regularization on them to balance the distribution of fitness value. Another positive effect of the regularization term is to accelerate the optimization process.

\subsection{Enhancing the edges}

\begin{figure}[tbp]
    \centering
    \begin{tabular}{c@{\hspace{1mm}}c@{\hspace{1mm}}c@{\hspace{1mm}}c@{\hspace{1mm}}c@{\hspace{1mm}}c}
        \includegraphics[height=1.5in]{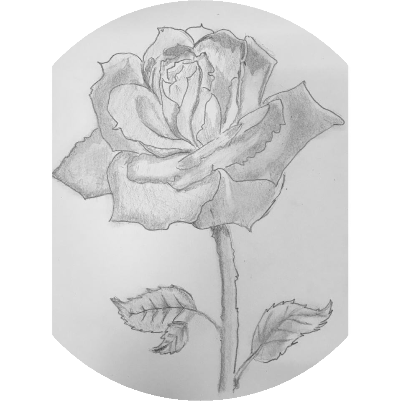}&
        \includegraphics[height=1.5in]{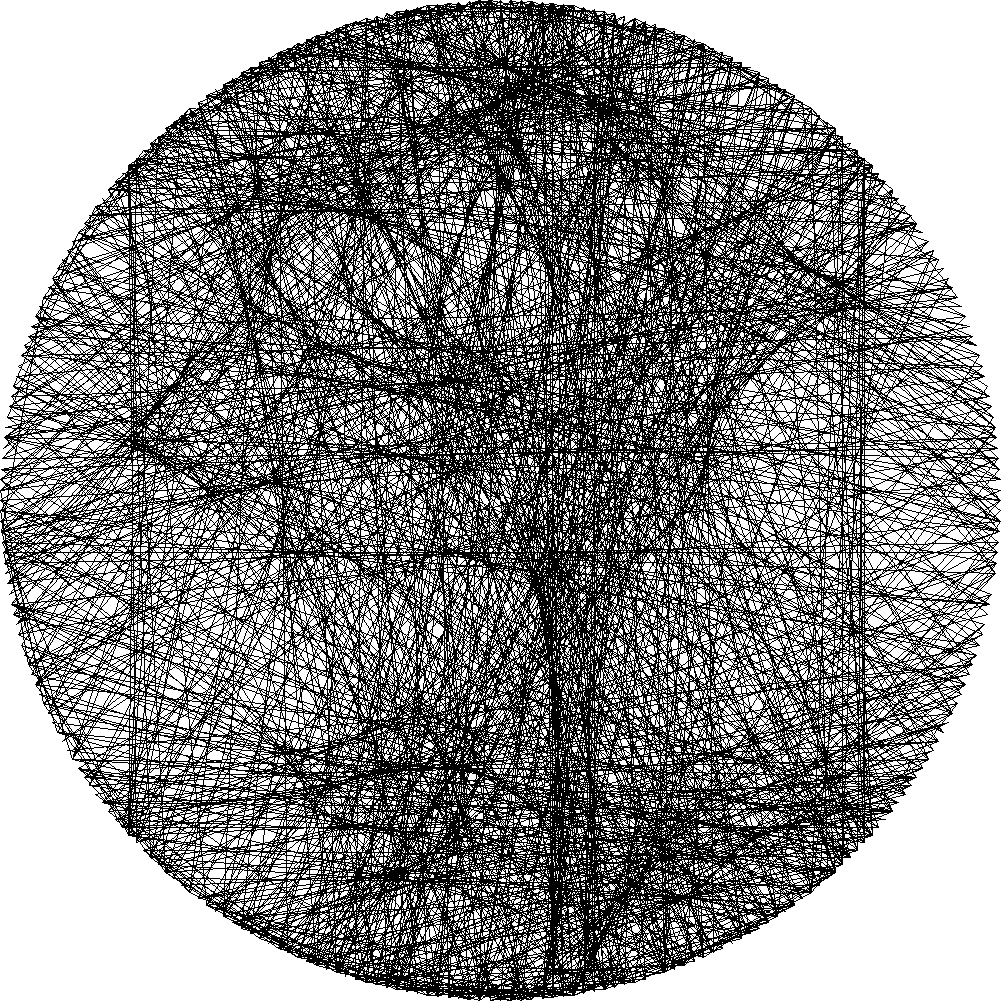}&
        \includegraphics[height=1.5in]{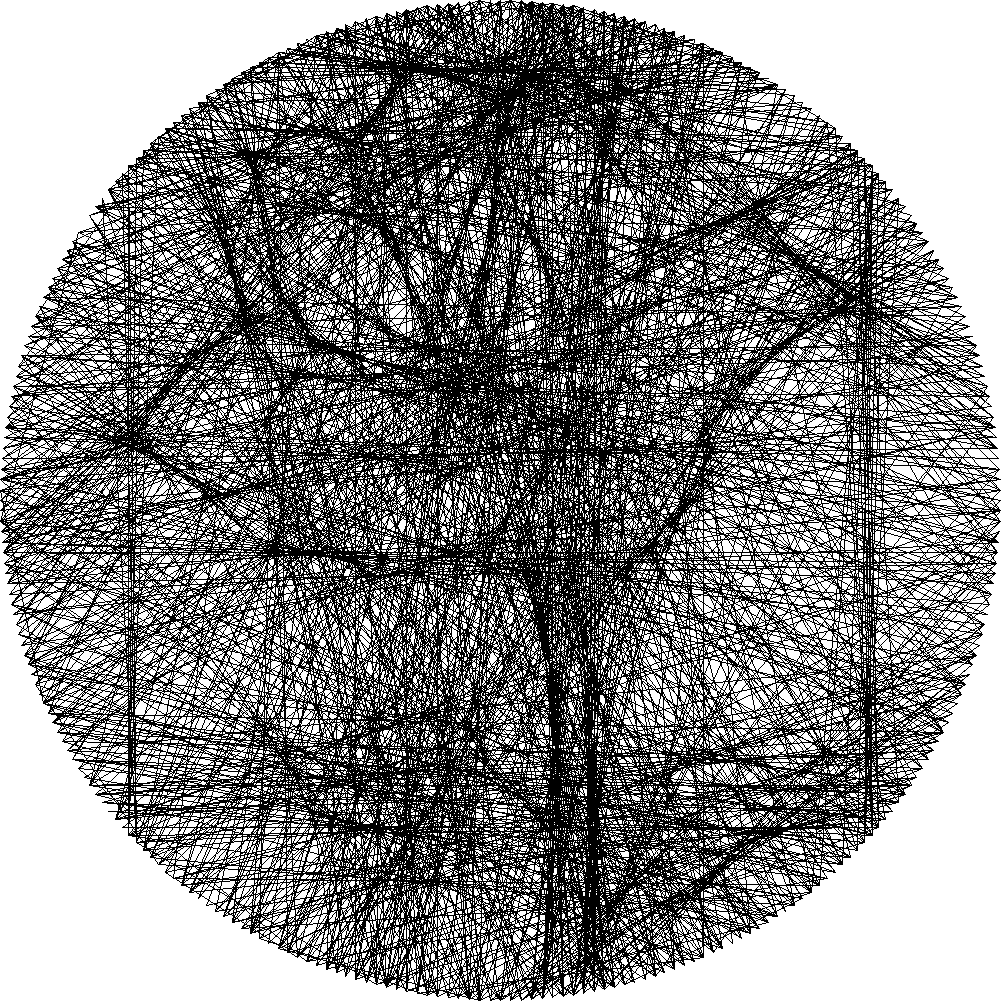}\\
                (a) & (b) & (c) \\
        \includegraphics[height=1.5in]{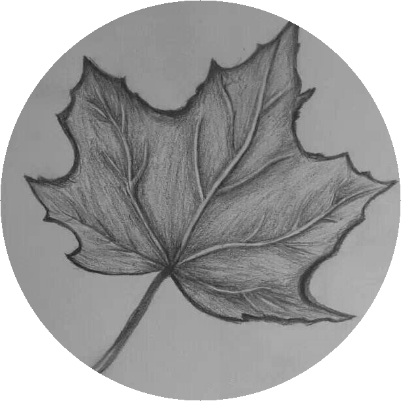}&
        \includegraphics[height=1.5in]{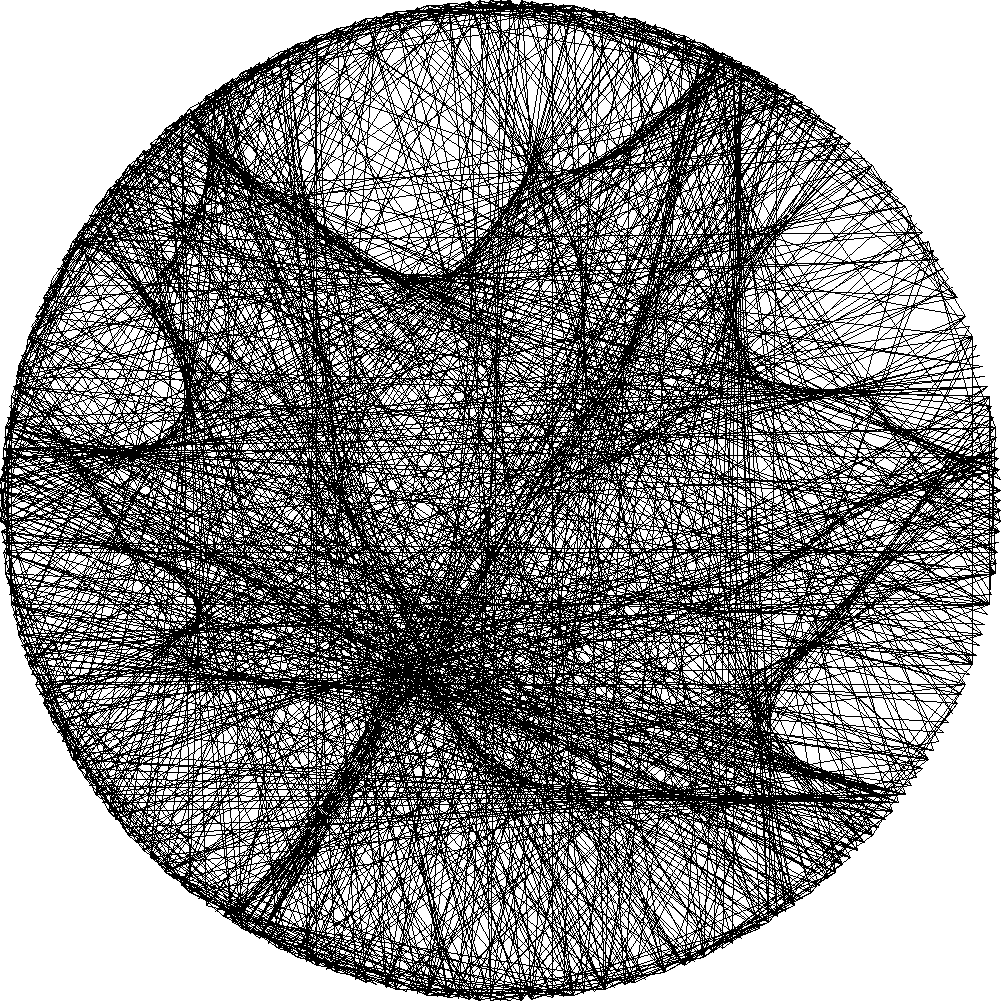}&
        \includegraphics[height=1.5in]{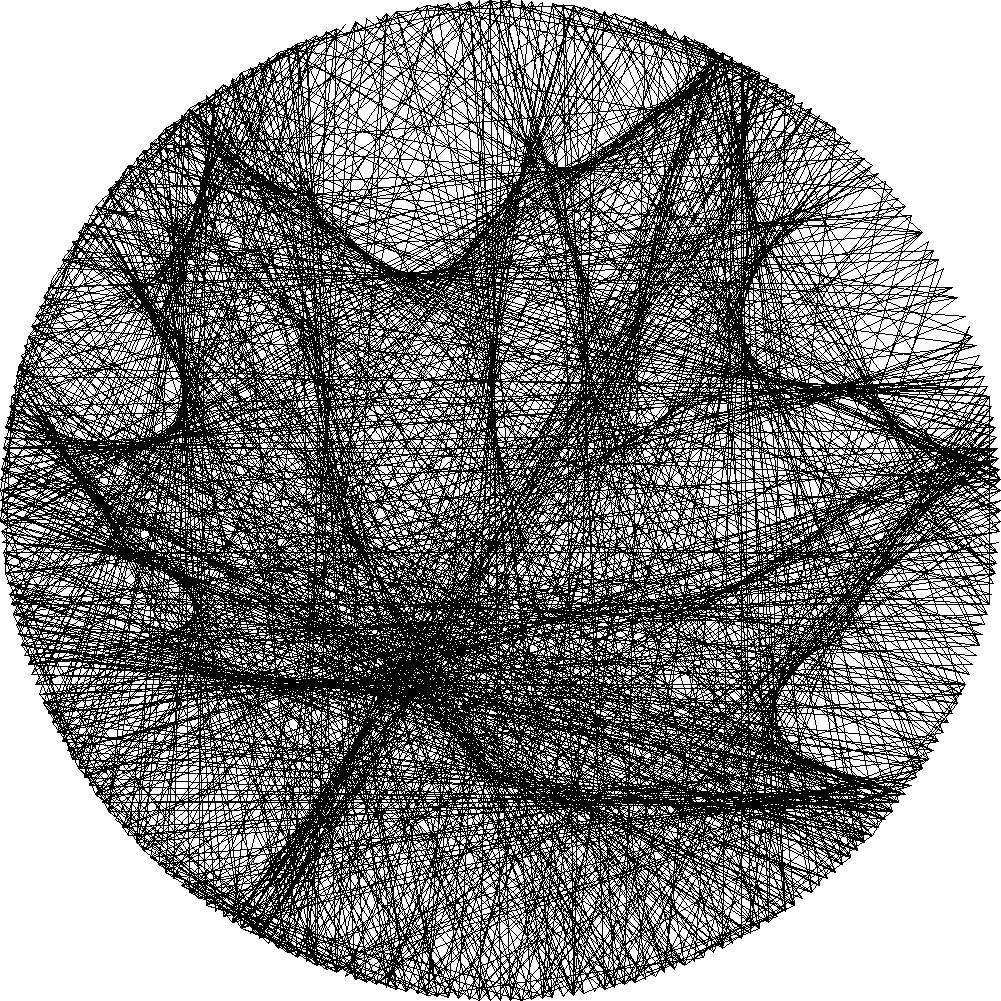}\\
        (d) & (e) & (f)  \\
    \end{tabular}
    \caption{The effect of edge enhancement by increasing the $\alpha$ parameter on the Flower (top) and Leaf (bottom) examples. (a) and (d) are the input images; (b) and (e) are the results without edge enhancement $(\alpha=0)$; (c) and (f) are the results with edge enhancement $(\alpha=0.5)$.}
    \label{fig:edge_enhance}
\end{figure}

In some images, the contrast between the main object and the background is not very distinct. Therefore, direct optimization using per-pixel gray-level value could not produce the clear outline of the object. To alleviate this problem, we increase the parameter $\alpha$ in Eq.\eqref{equ_edge} from $0$ in previous experiments to $0.5$. Fig.~\ref{fig:edge_enhance} illustrates two examples with edge-enhancement.  Note that in these cases the overall similarity measure between the inputs and the outputs was decreased, but the clarity of the objects was increased, and  better visual quality was achieved.

\begin{figure}[tbp]
    \centering
    \begin{tabular}{c@{\hspace{4mm}}c}
        &
        \begin{tabular}{c@{\hspace{1mm}}c@{\hspace{1mm}}c@{\hspace{1mm}}c@{\hspace{1mm}}c}
            \includegraphics[height=1.1in]{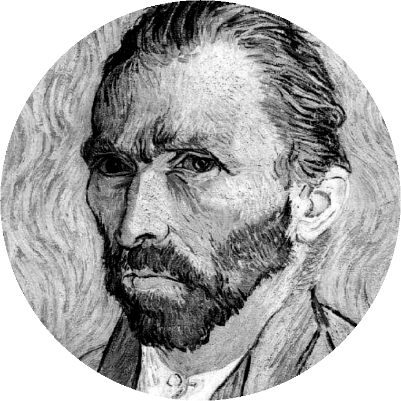}&
            \includegraphics[height=1.1in]{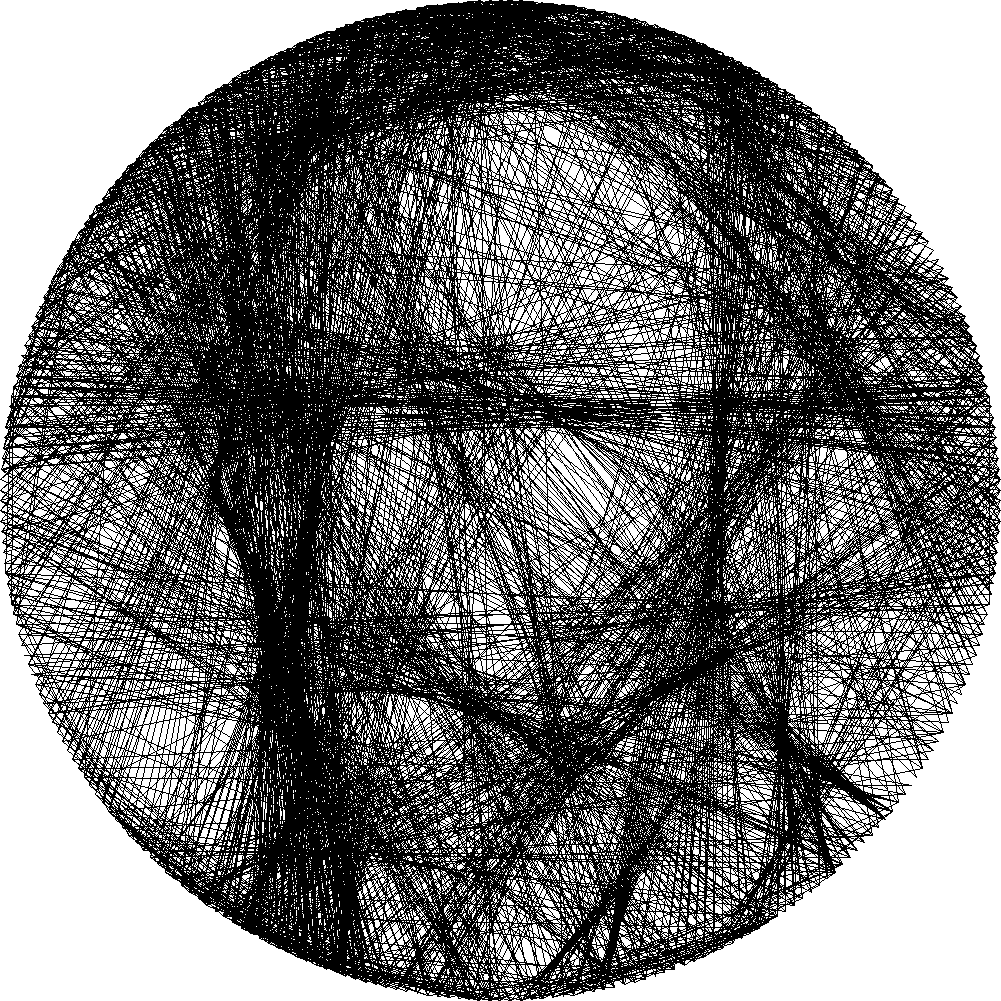}&
            \includegraphics[height=1.1in]{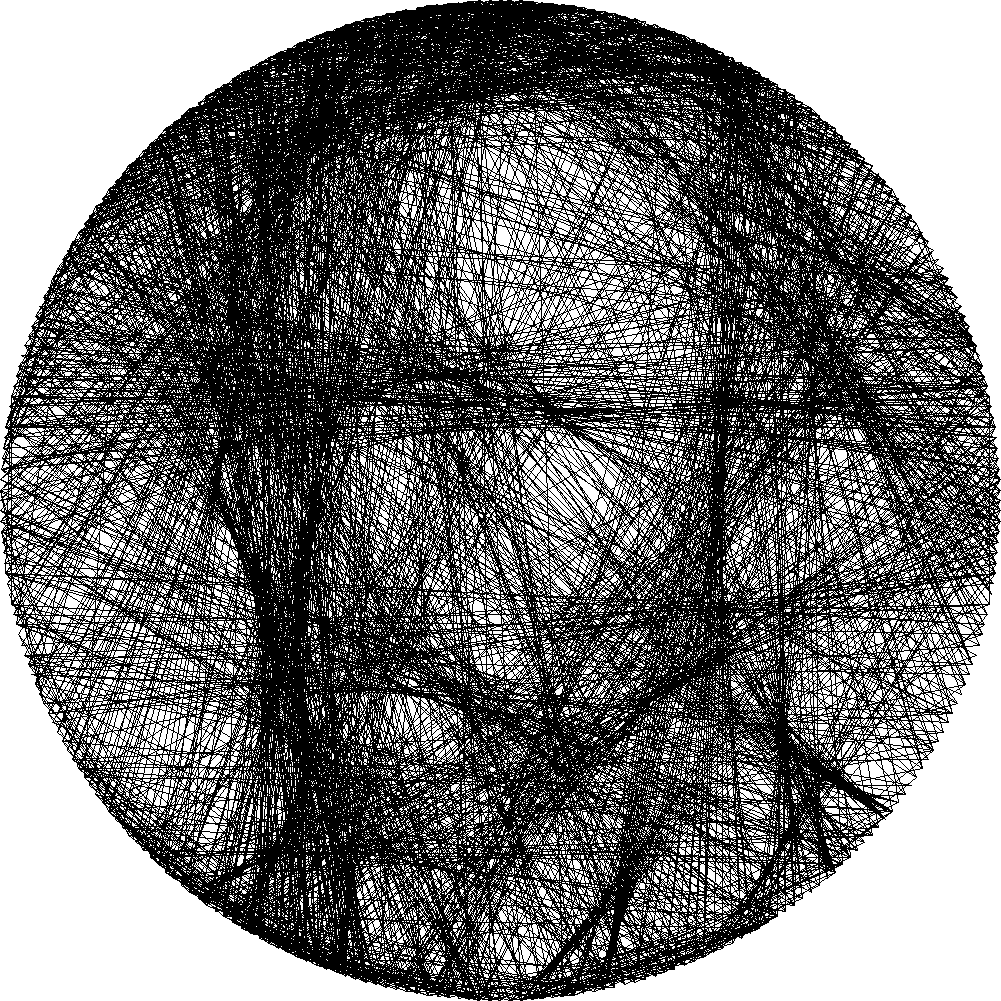}&
            \includegraphics[height=1.1in]{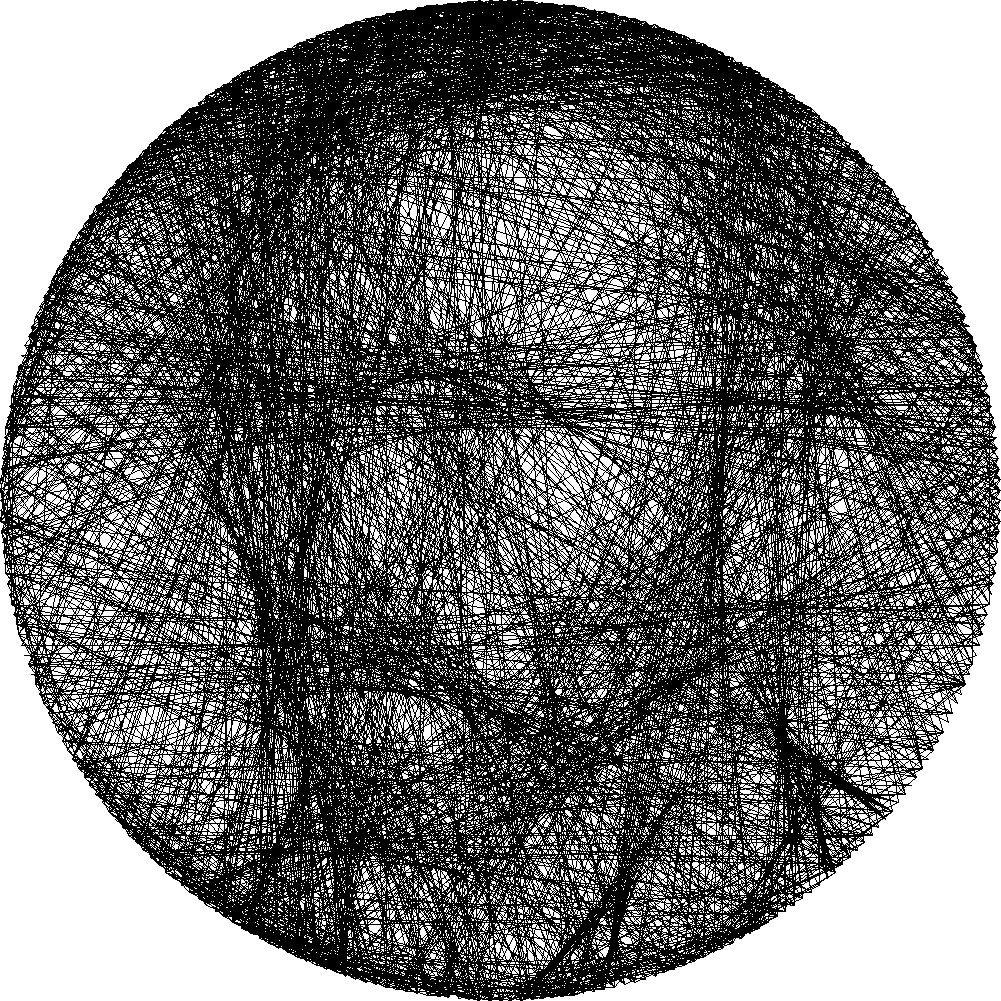}\\
            (a) & (b) & (c) & (d)
        \end{tabular}
    \end{tabular}
    \caption{The effect of changing the sharpness factor $T$ on the Van Gogh example. (a) original image; (b)-(d) outputs corresponding to varying $T=10,20,40$.}
    \label{fig:sharpness}
\end{figure}

\begin{figure}[tbp]
    \centering
    \begin{tabular}{c@{\hspace{4mm}}c}
        &
        \begin{tabular}{c@{\hspace{1mm}}c@{\hspace{1mm}}c@{\hspace{1mm}}c@{\hspace{1mm}}c}
            \includegraphics[height=1.1in]{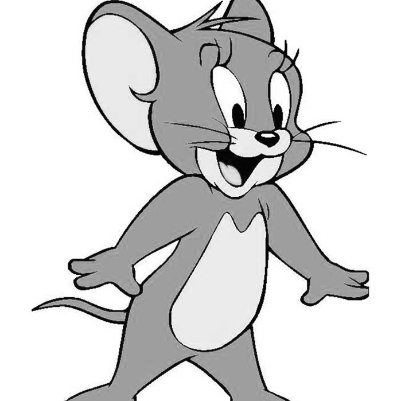}&
            \includegraphics[height=1.1in]{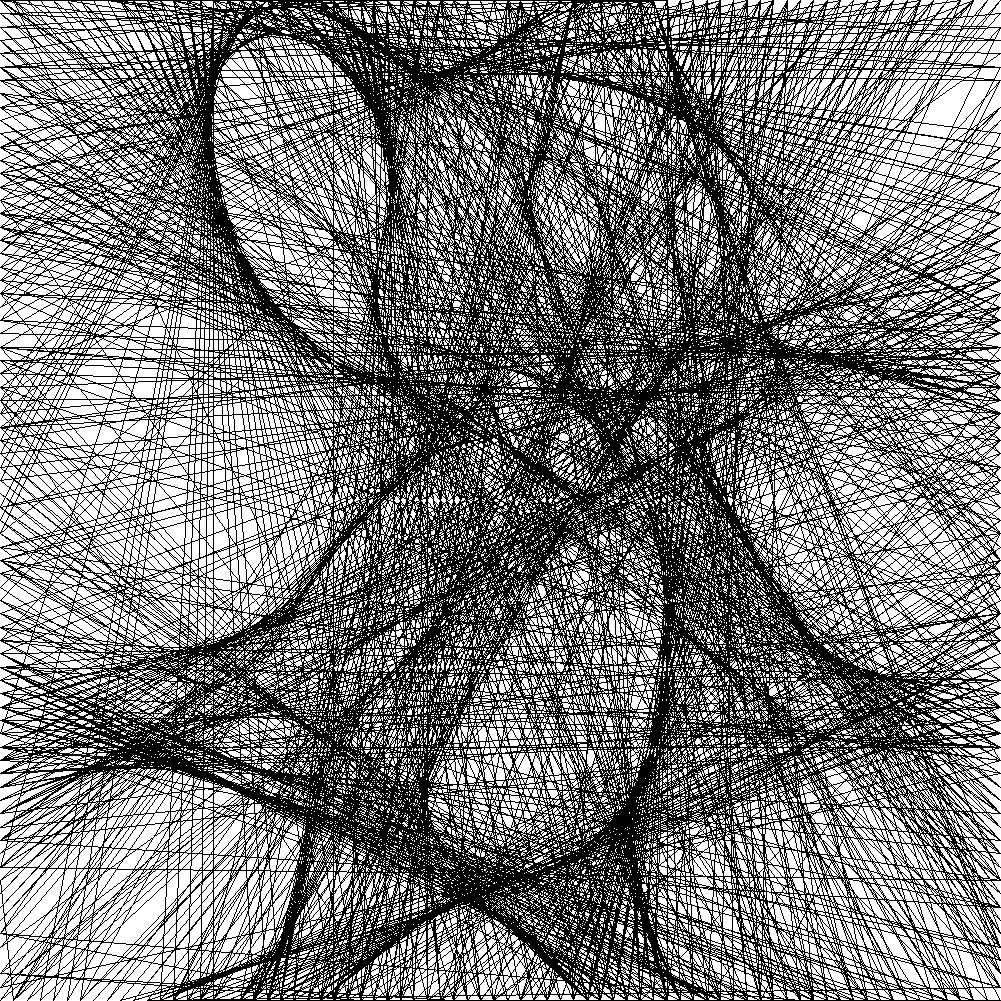}&
            \includegraphics[height=1.1in]{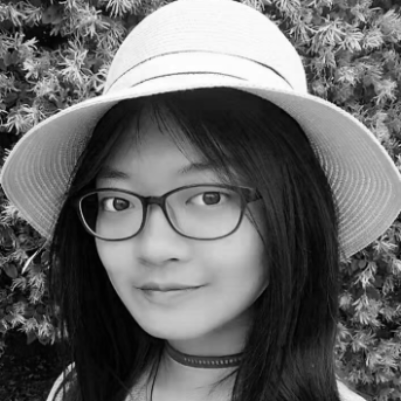}&
            \includegraphics[height=1.1in]{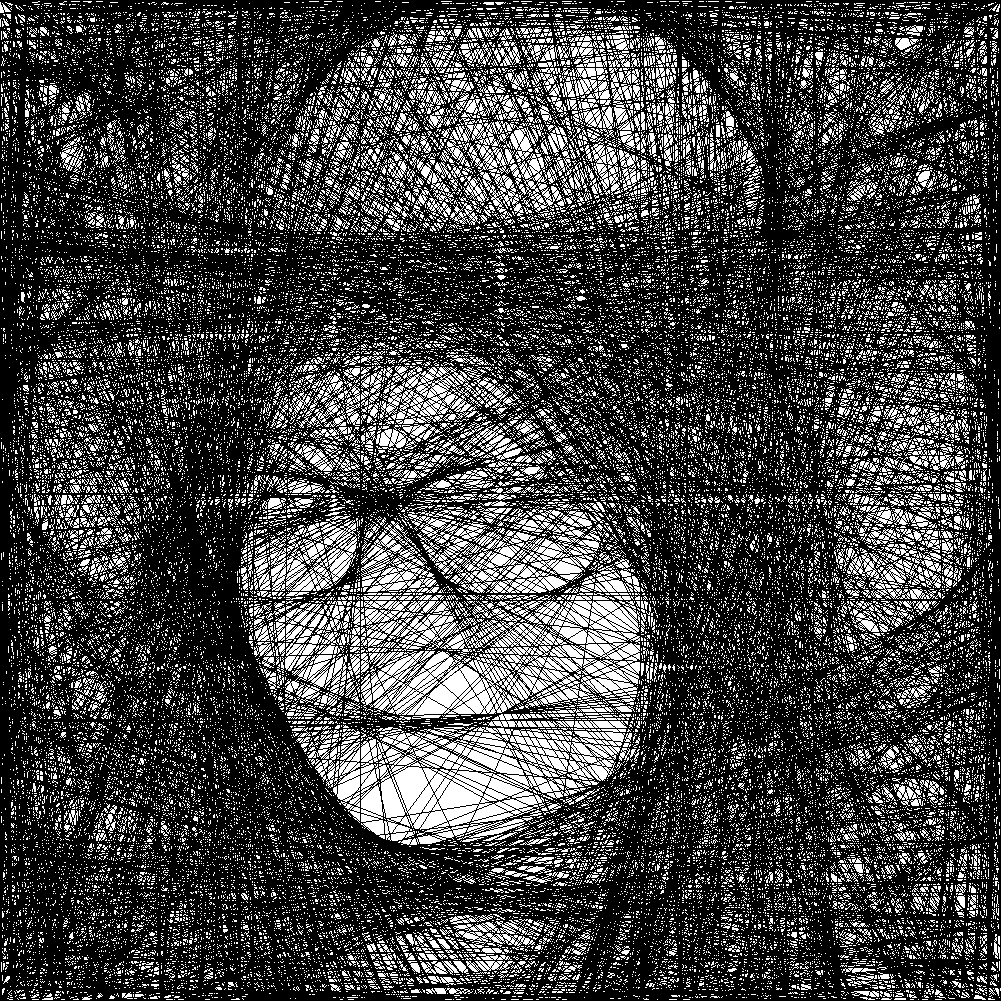}\\
            (a) & (b) & (c) & (d)
        \end{tabular}
    \end{tabular}
    \caption{Examples of square thread paintings. (a), (c): Input image; (b), (d): Square painting.}
    \label{fig:square}
\end{figure}

\subsection{Variation of sharpness}

The parameter $T$ controls the amount of diffused error and hence sparsity of thread painting. Fig.\ref{fig:sharpness} illustrates the trend of outputs when increasing $T$. As $T$ is increases, the chords become more separated. An appropriate value of $T$ leads to having both  distinct edges and a good reconstruction of the whole image.


\subsection{Changing the border shape}
In the examples above, as well as the original work of Petros Vrellis, the shape of the canvas is a circle. However, there is no reason to constrain the boundary shape to one shape.  In fact, we can use other boundary shapes, such as squares, to generate various results (see Fig. \ref{fig:square}). In this case, because some pins lie on a straight line, we exclude the chords lying on these lines from the optimization process as they do not contribute to the painting.  

\begin{table}[tbp]
        \begin{center}
        \resizebox{\textwidth}{!}{\begin{tabular}{l||c|c|c||c|c}
\hline
& \multicolumn{3}{c|}{average rank} & & \\
\cline{2-4}
\multirow{2}{*}{Image Name} & \multirow{2}{*}{Greedy} & \multirow{2}{*}{Ours} & Ours & average & Expected number\\
         &       &     & (disconnected)        &     gray level     & of chords\\
\hline
Jerry            & 2.95 & \textbf{1.43} & 1.62 & 205 & 1250 \\
Winnie           & 3.00 & 1.52 & \textbf{1.48} & 217 & 1155 \\
Girl 1           & 2.48 & \textbf{1.29} & 2.24 & 110 & 1857 \\
Girl 2           & 2.81 & \textbf{1.24} & 1.95 & 109 & 1869 \\
Poetin           & 2.33 & \textbf{1.38} & 2.29 & 157 & 1607 \\
Trump            & 2.86 & \textbf{1.24} & 1.90 & 191 & 1607 \\
Van Gugh         & 2.62 & 1.86 & \textbf{1.52} & 146 & 1583 \\
Du Fu            & 2.76 & \textbf{1.14} & 2.10 & 173 & 1595 \\
Leaf             & 2.86 & \textbf{1.14} & 2.00 & 134 & 1285 \\
Flower           & 2.95 & \textbf{1.33} & 1.71 & 208 & 1202 \\
Nuclear          & 2.05 & \textbf{1.52} & 2.43 &  76 & 1595 \\
Leonardo         & 2.81 & \textbf{1.29} & 1.90 & 134 & 1667 \\
Jobs             & 2.86 & \textbf{1.38} & 1.76 & 188 & 1345 \\
Mario            & 3.00 & \textbf{1.48} & 1.53 & 217 & 1297 \\
Mushroom         & \textbf{1.52} & 2.43 & 2.04 & 175 & 1690 \\
\hline
        \end{tabular}}
\end{center}
\caption{Columns 2-4: Average user labelled rank (1 is best, 3 is worst) for the three different methods. Columns 5-6 compare the average gray level of input image to the average number of cords for best ranked ThreadTone painting. The correlation is around $-0.7$, meaning that the darker the image is, the more chords are needed.}
\label{tab:user_study}
\end{table}

\subsection{User study}
\label{section:userstudy}
We invited 20 student volunteers to evaluate the quality of the thread paintings. The original images, as well as the results of three approaches (greedy, ours, and ours without connectivity requirement) were provided to the evaluators, who were required to rank the quality of the three paintings. The results of the study are shown in Table~\ref{tab:user_study}. Overall, our method with connectivity requirement was regarded as the best one.

Moreover, we presented paintings produced by our method with different number of chords (750, 1000, 1250, 1500, 1750, 2000) and then asked the evaluators to select the best one. Result given in Table~\ref{tab:user_study} indicates that, in general, the appropriate amount of chords is proportional to the darkness of the input image. However, the complexity of describing the structure differs from one image to the other, and there is no simple rule that can suit all the different cases.

\section{Conclusion}

In this work, we formulated the thread painting problem and provided an automatic solution to produce such painting from input images. The proposed algorithm consists of two parts. First, we compute the fitness function of the chord space by solving a least square minimization problem. The objective function is a combination of per-pixel reconstruction loss and a regularization term on the chord fitness value. Different weights are assigned to emphasize the quality of important regions, and to provide preference to chords with suitable length and direction. After acquiring the fitness values of chords, a sampling process is conducted to form a sequence of connected chords to be drawn in the circle. Error diffusion is applied in the neighbourhood of selected chords during sampling to control the sharpness of the result. We evaluated the thread paintings results with SSIM index and a user study. Results show that our approach can create thread paintings with high quality on various inputs.

\address{Department of Computer Science and Technology\\
 Tsinghua University, Beijing, China\\
\email{wwjpromise@163.com}}

\address{Department of Computer Science and Technology\\
 Tsinghua University, Beijing, China\\
\email{bin-liu13@mails.tsinghua.edu.cn}}

\address{Efi Arazi School of Computer Science\\
the Interdisciplinary Center, Herzliya\\
\email{arik@idc.ac.il}}

\end{document}